\documentclass[journal]{IEEEtran}

\IEEEoverridecommandlockouts                              

\usepackage{lastpage}
\usepackage{fancyhdr}

\usepackage[pdftex]{graphicx}
\graphicspath{{figure/}}
\usepackage{amsmath}
\usepackage{amssymb}
\usepackage{url}

\usepackage{amsthm} 
\usepackage{cite}
\usepackage{bm} 

\usepackage{algorithm}
\usepackage{algorithmic}
\usepackage{amssymb}
\usepackage{color}

\newcommand{\mc}{\mathcal}
\newcommand{\mb}{\mathbf}
\newcommand{\C}{\mathbb{C}}

\newcommand{\bx}{\mathbf{x}}
\newcommand{\bv}{\mathbf{v}}
\newcommand{\bi}{\mathbf{i}}
\newtheorem{proposition}{\textbf{Proposition}}

\newboolean{showcomments}
\setboolean{showcomments}{true}
\newcommand{\lina}[1]{  \ifthenelse{\boolean{showcomments}}
	{ \textcolor{red}{(Lina says:  #1)}} {}  }

\usepackage{multirow}

\allowdisplaybreaks

\begin{document}

	\title{Aggregate Power Flexibility in Unbalanced Distribution Systems}


\author{Xin Chen, Emiliano Dall'Anese, 
	Changhong Zhao, Na Li
	\thanks{  X. Chen and N. Li are with the School of Engineering and Applied Sciences, Harvard University, USA. Email: (chen\_xin@g.harvard.edu, nali@seas.harvard.edu). E. Dall'Anese is with the Department of Electrical, Computer, and Energy Engineering, 
		University of Colorado Boulder, USA. Email: 
		emiliano.dallanese@colorado.edu. C. Zhao is with the Department of Information Engineering, the Chinese University of Hong Kong, China. Email: zhchangh1987@gmail.com.}
	\thanks{ 
		The work was supported by NSF 1839632,
		 NSF CAREER 1553407, ARPA-E through the NODES program, and Harvard Climate Change Solution Funds.
	} 
}



\maketitle
	\pagestyle{plain}

\thispagestyle{plain}          


\begin{abstract}

With a large-scale integration of distributed energy resources (DERs), distribution systems are expected to be capable of providing capacity support for the transmission grid. To effectively harness the collective flexibility from massive DER devices, this paper studies distribution-level power aggregation strategies for transmission-distribution interaction. In particular, this paper proposes a method to model and quantify the aggregate power flexibility, i.e.,  the net power injection achievable at the substation, in unbalanced distribution systems over time. Incorporating the network constraints and multi-phase unbalanced modeling, the proposed method obtains an effective approximate feasible region of the net power injection. For any aggregate power trajectory within this region, it is proved that there exists a feasible disaggregation solution. In addition, a distributed model predictive control (MPC) framework is developed for practical implementation of the transmission-distribution interaction.
At last, we demonstrate the performances of the proposed method via numerical tests on a real-world distribution feeder with 126 multi-phase nodes.

\end{abstract}

	
		\section*{Nomenclature}

\addcontentsline{toc}{section}{Nomenclature}

{ 

\subsection{Parameters}
\begin{IEEEdescription}[\IEEEusemathlabelsep\IEEEsetlabelwidth{$V_1,V_2,V_3$}]
	\item [$N$] Number of buses (except the substation).
	\item [$N_Y,\,N_\Delta$] Number of buses with wye-, delta-connection.
	\item [$L$] Number of distribution lines.
	\item [$\bar{\mb{v}},\,\underline{\mb{v}}$]  Upper, lower limits of the three-phase nodal voltage magnitudes for all buses. 
	\item [$\bar{\mb{i}}_L,\, \underline{\mb{i}}_L $] Upper,  lower limits of the three-phase line current magnitudes for all distribution lines.
	\item [$\bar{P}_{i,t}^{g,\psi},\underline{P}_{i,t}^{g,\psi}$] Upper, lower 
	limits of active PV power generation in phase $\psi$ of bus $i$ at time $t$.
	\item [$\bar{S}_{i,t}^{g,\psi}$] 
    Apparent power capacity of PV units in phase $\psi$ of bus $i$ at time $t$.
    \item [$\bar{P}_{i,t}^{e,\psi},\underline{P}_{i,t}^{e,\psi}$]  Upper, lower 
    limits of active power output of ES devices in phase $\psi$ of bus $i$ at time $t$.
    \item [$\bar{S}_{i,t}^{e,\psi}$] Apparent power capacity of ES devices in phase $\psi$ of bus $i$ at time $t$.
    \item [$\bar{E}_i, \underline{E}_i$] 
     Upper, lower limits for state of charge of  ES devices at bus $i$.
     \item [$\bar{P}_{i,t}^{d,\psi},\underline{P}_{i,t}^{d,\psi}$]  Upper, lower limits for  controllable active loads in phase $\psi$ of bus $i$ at time $t$.
     \item [$F_{i,t}^{out}$] Outside temperature for HVAC systems
      at bus $i$ at time $t$.
     \item [$\bar{F}_{i},\underline{F}_{i}$] Upper, lower limits of comfortable temperature zone for HVAC systems
     at bus $i$.
     \item [$\Delta t$]  Length of each time slot under discretized time horizon.
\end{IEEEdescription}	

\subsection{Variables}
	\begin{IEEEdescription}[\IEEEusemathlabelsep\IEEEsetlabelwidth{$V_1,V_2,V_3$}]
		 \item[\smash{\begin{IEEEeqnarraybox*}[][t]{l}
		 \mb{p}_Y, \mb{q}_Y\\
	 	\hphantom{}\in \mathbb{R}^{3N_Y}
		\end{IEEEeqnarraybox*}}] Column vector of the three-phase   active,
		{reactive power injection  via  wye-connection.}
        \item[\smash{\begin{IEEEeqnarraybox*}[][t]{l}
		\mb{p}_\Delta, \mb{q}_\Delta\\
		\hphantom{}\in \mathbb{R}^{3N_\Delta}
        \end{IEEEeqnarraybox*}}] Column vector of the three-phase   active,
        {reactive  power injection  via  delta-connection.}
        	\item[$\mb{v}\in\mathbb{R}^{3N}$] Column vector collecting the three-phase nodal voltage magnitudes for all buses.
        \item[$\mb{i}_L\in \mathbb{R}^{3L}$] Column vector collecting the three-phase  line current magnitudes for all distribution lines.
        \item[$\mb{p}_{0}\in\mathbb{R}^3$] Column vector of the three-phase net active  power injection at the substation. 
       \item [$P_{i,t}^{g,\psi},Q_{i,t}^{g,\psi}$] Active, reactive PV  power generation in phase $\psi$ of bus $i$ at time $t$.
       \item [$P_{i,t}^{e,\psi},Q_{i,t}^{e,\psi}$] Active, reactive power output of ES devices in phase $\psi$ of bus $i$ at time $t$.
       \item [$E_{i,t}$] State of charge of ES devices at bus $i$ at time $t$.
       \item [$P_{i,t}^{d,\psi},Q_{i,t}^{d,\psi}$] Active, reactive controlled loads in phase $\psi$ of bus $i$ at time $t$.
       \item [$P_{i,t}^{h,\psi},Q_{i,t}^{h,\psi}$]  Active, reactive HVAC loads in phase $\psi$ of bus $i$ at time $t$.
       \item [$F_{i,t}^{in}$] Indoor temperature for HVAC systems
       at bus $i$ at time $t$.
       \item [$P_{0,t}$] Total net active power injection (summation over three phases) at the substation at time $t$.
\end{IEEEdescription}

Note: the same notations without  superscript $\psi$ denote the corresponding summation over phases, e.g. $P_{i,t}^{e}: = \sum_{\psi}P_{i,t}^{e,\psi} $.

\subsection{Notation}
\begin{IEEEdescription}[\IEEEusemathlabelsep\IEEEsetlabelwidth{$\jmath:=\sqrt{-1}$}]
	\item[$\jmath$] $:=\sqrt{-1}$, i.e., the imaginary unit.
		\item[$\mathbf{1}_n$] $n\times1$ column vector with all ones. 	
		\item[$|\cdot|$] Entry-wise absolute value of a vector.
	\item[$\text{diag}(\mb{x})$]  Diagonal matrix with the elements of vector $\mb{x}$ in its diagonal.
	\item[$\mathcal{R}\{\cdot\}$] Real part of a complex value.
	\item[$(\cdot)^*$] Complex conjugate.
	\item[$(\cdot)^\top,(\cdot)^{-1}$] Matrix transposition and matrix inverse.
	\item[$\frac{\partial f(x)}{\partial x}$] Partial derivative. 
\end{IEEEdescription}

}
	
\section{Introduction}


\IEEEPARstart{C}{onventionally}, a distribution grid is treated as an equivalent passive load in the operation of transmission systems due to its non-dispatchability \cite{m-s}. In the past decade, a rapid proliferation of distributed energy resources (DERs) has been witnessed in distribution systems, especially photovoltaic (PV) generation, energy storage (ES) and demand response. 
As the penetration of dispatchable DERs deepens, 
 significant flexibility has been introduced to electricity distribution, evolving the distribution systems from \emph{passive} to \emph{active} state \cite{ADN}. 
 {Similar to the transmission systems, the  power flexibility of the distribution systems refers to the extra power capacity that can be dispatched to maintain safe and efficient system operation especially in case of contingency. 
 While distinguished from the large-scale
spinning and non-spinning reserves in the transmission systems,  the distribution flexibility is provided by a large amount of heterogeneous DERs. Although each DER usually has a small capacity, the coordinated dispatch of ubiquitous DERs is capable of providing significant flexibility support and therefore enables active participation of distribution systems in the transmission system operation.}
 Through coordinated transmission-distribution dispatch, the power grids can fully exploit the potential flexibility and achieve greater efficiency and resilience.

\begin{figure}[thpb]
	\centering
	\includegraphics[scale=0.5]{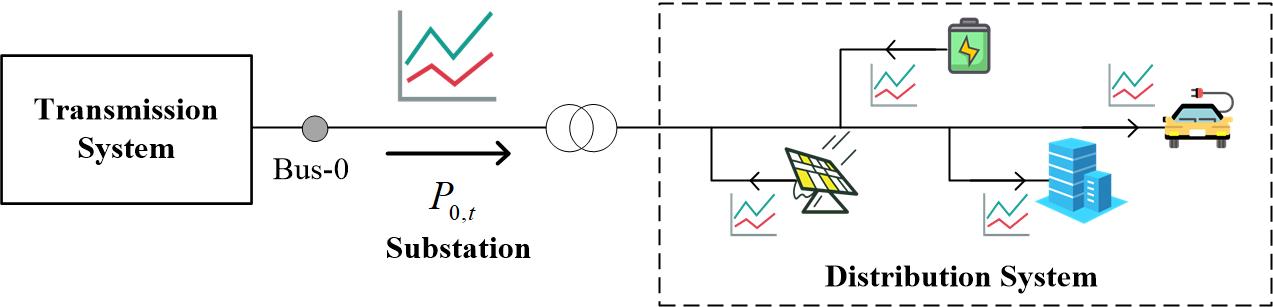}
	\caption{Illustration of power aggregation in distribution networks.}
	\label{f_c2}
\end{figure}

In practice, managing a large population of DER devices for system-wide operation and control is challenging due to the computational complexity. {References\cite{td-1,td-2,td-3} apply decomposition methods to incorporate transmission and distribution in economic dispatch and reactive power optimization. However, these methods require a number of iterations and boundary information exchanges between transmission and distribution, and may suffer from slow convergence issues given
 the sheer scale of this problem due to the large amount of DERs,
thus they are arduous for practical applications.} To address this challenge, as a promising alternative, power aggregation has received considerable attention for leveraging the available flexibility from the distribution side. As illustrated in Figure \ref{f_c2}, the generation or power consumption of each DER device can be captured by a certain feasible region, which is specified by its own operational constraints and dynamics. Power aggregation is to model and qualify the aggregate flexibility at the  substation, which is the achievable  net power injection to the distribution feeder.
{By reporting this concise and compact feasible region, the distribution grid can actively participate in the operation and control of
transmission systems as a \textit{virtual power plant} \cite{vpp}}. Hence, this paper focuses on developing novel power aggregation methods for distribution systems, which takes into account both the network model and DER device models.

Basically, power aggregation can be regarded as a projection of the high-dimensional operational constraints onto the feasible region of the net substation load. However, considering tens of thousands of electric devices and multiple time steps, procuring this projection is computationally intensive and impractical.
Hence most research efforts are devoted to building inner or outer approximations of the exact feasible region. Reference \cite{eli} utilizes a series of time-moving ellipsoids to model the aggregate P-Q feasible domain over time, and follows a data-driven system identification procedure to obtain the model parameters. In \cite{min_1,min_2,min_3}, individual flexibility of each DER is described as a polytopic feasible set, then the aggregate flexibility is calculated as the Minkowski sum of the individual polytopes. Reference \cite{pro2} applies the polytopic projection to procure the aggregate flexibility and formulates an approximated optimization problem for tractable solution. In \cite{opf_1} \cite{opf_2}, robust optimization models are developed for estimating and optimally scheduling the aggregate reserve capacities considering the uncertainties of regulation signals and forecast errors.

Many researches above focus on a single type of DERs, e.g., thermostatically controlled loads (TCL) or heating, ventilation and air-conditioning (HVAC) systems. Then their flexibility modeling and aggregation approaches may not be able to handle a variety of DERs in distribution systems.
 In addition, most existing studies \emph{disregard} the network constraints, such as  voltage limits and line thermal constraints,  which are crucial to the system-level power aggregation. Furthermore, since distribution networks are intrinsically unbalanced due to non-symmetrical conductors, untransposed lines, and unequal interphase power injection \cite{unb}, 
the multi-phase modeling of networks and DER devices is necessary.


To address these challenges, this paper proposes an power aggregation method to quantify the aggregate flexibility from different types of DERs in unbalanced distribution systems. In particular, we approximate the exact feasible region of the net power injection at the substation with an inner-box region, and the upper and lower operational trajectories are defined to specify the aggregate flexibility over time. Then two multi-period optimization models are established for system-level power aggregation: one aims to evaluate the maximal flexibility level of the distribution systems, and the other is to co-optimize the base operational trajectory and flexibility reserve capacities. The main contributions of this paper are summarized as follows.

\begin{itemize}
	\item  [(1)] A distribution-level power aggregation method is proposed, which incorporates both the network constraints and multi-phase unbalanced modeling. The existence of power disaggregation solution is guaranteed with our method (see Proposition \ref{pro1}).
	
	\item  [(2)] To protect the privacy of DER facilities and enable scalable application, we develop a distributed model predictive control (MPC) framework for practical implementation of the distribution-transmission interaction.
\end{itemize}

{Our proposed framework can be viewed as a hierarchical structure for the distribution-transmission interaction: each distribution system performs its own power aggregation, and the transmission system manages the coordination between generation and various distribution systems. As a result, the distribution-transmission interaction can be achieved in an easy and efficient manner, which is promising for practical implementation.}

The remainder of this paper is organized as follows: Section II introduces the multi-phase network and DER models. Section III presents the inner-box approximation method and  two power aggregation  models. Section IV 
develops the distributed MPC framework for transmission-distribution interaction. Numerical tests are carried out on a real feeder system in Section V, and conclusions are drawn in Section VI.

\section{Network and DER Models}



In this section, we consider the multi-phase wye- and delta-connection of electric devices, and 
present the  network model and DER models in unbalanced distribution systems.

\subsection{Network Model}
\begin{figure}[thpb]
	\centering
	\includegraphics[scale=0.35]{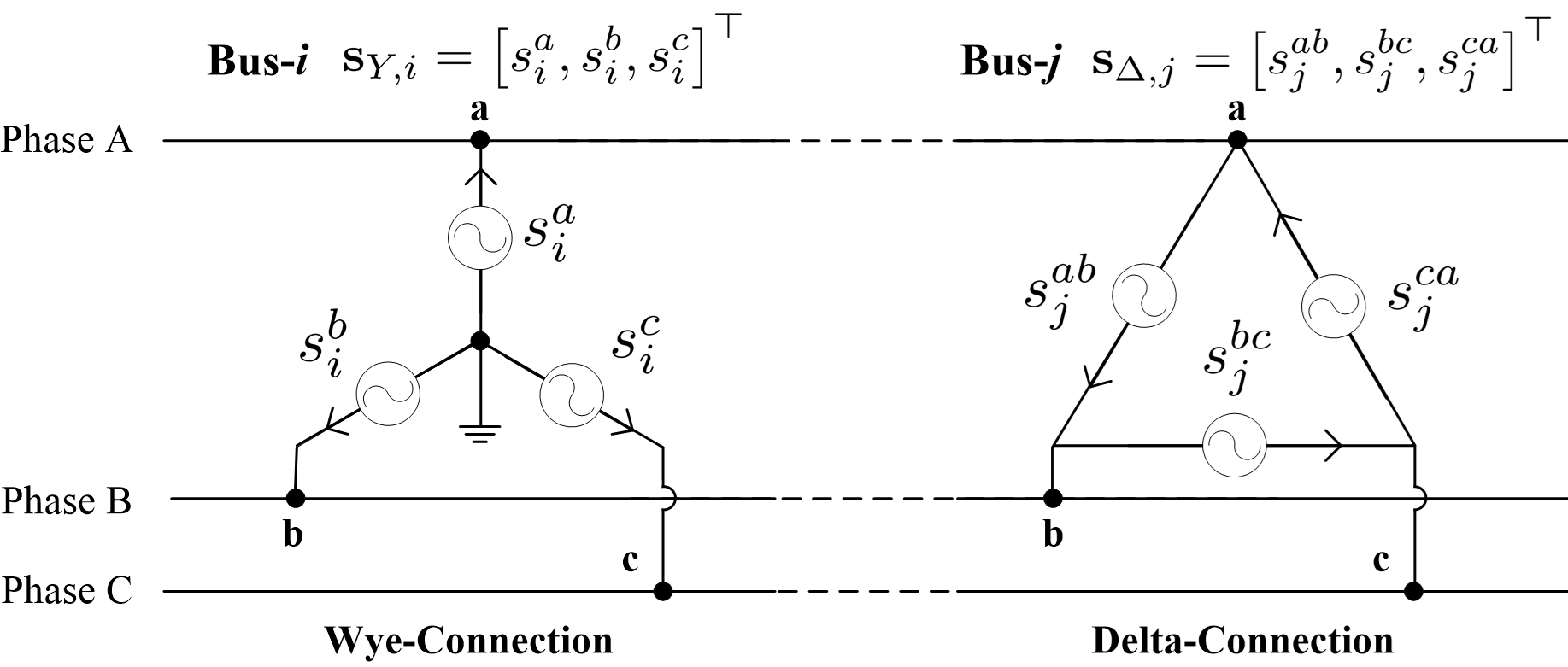}
	\caption{Illustration of wye-connection and delta-connection.}
	\label{w_d}
\end{figure}

Consider a multi-phase distribution network described by the graph $G({\mc{N}}_0,\mc{E})$, where $\mc{N}_0$ denotes the set of buses and $\mc{E}\subset \mc{N}_0\times \mc{N}_0$ denotes the set of distribution lines. Let $\mc{N}_0:=\mc{N} \cup \{0\}$ with $\mc{N}:=\{1,2,\dots,N\}$ and bus-$0$ is the slack bus, i.e., the substation interface to the transmission grid. 
As shown in Figure \ref{w_d}, each electric device can be multi-phase wye-connected or delta-connected  to the distribution network \cite{wye}. Denote  $\phi_{Y}:=\left\{a,b,c\right\}$ and $\phi_{\Delta}:=\left\{ab,bc,ca\right\}$. Then we use notation $\psi$ to describe the concrete connection manner of an electric device  with either $\psi\subseteq \phi_{Y}$ or $\psi\subseteq \phi_{\Delta}$. 
For instance,  $\psi=\left\{a\right\}$ if the device is wye-connected in phase A and only has the complex power injection $s^a$, while $\psi=\left\{ab,bc\right\}$ if it is delta-connected in phase AB and BC with the complex power injection $s^{ab}$ and $s^{bc}$. 
Denote $\mathcal{N}_Y$ and $\mathcal{N}_\Delta$ as the set of buses with wye- and delta-connection respectively.
Let column vector $\mb{s}_Y:=\mb{p}_Y+\jmath\,\mb{q}_Y\in \mathbb{C}^{3N_Y}$ and $\mb{s}_\Delta:=\mb{p}_\Delta+\jmath\,\mb{q}_\Delta\in\mathbb{C}^{3N_\Delta}$ collect all the three-phase complex power injection via wye- and delta-connection respectively. 


{{For compact expression,  we stack all the three-phase power injection  into a long real-value vector as}
\begin{align}
\mb{x}:=\left[\mb{p}_Y^\top,\mb{q}_Y^\top,\mb{p}_\Delta^\top,\mb{q}_\Delta^\top \right]^\top  \label{dex}
\end{align}}
With the fixed-point linearization method introduced in \cite{linear_1},
we can derive the linear multi-phase power flow model (\ref{lp1})
with a given operational point.
\begin{subequations} \label{lp1}
	\begin{align}
	\mb{v}\ \, &=\mb{A}\mb{x}+\mb{a} \label{v}\\
	\mb{i}_L\, &=\mb{B}\mb{x}+\mb{b} \label{il}\\
	\mb{p}_{0}&=\mb{D}\mb{x}+\mb{d} \label{p_0}
	\end{align}
\end{subequations}
Here, matrices $\mb{A}, \mb{B}, \mb{D}$ and vectors $\mb{a},\mb{b},\mb{d}$ are system parameters, whose definitions are briefly provided in Appendix A. See \cite{linear_2} for more details on the linear power flow model (\ref{lp1}).

{  Specifically, the fixed-point linearization method \cite{linear_1} 
reformulates the exact power flow equations as a fixed-point form, then the linear model (\ref{lp1}) is obtained by running a single iteration of the fixed-point equations at the given  operational point. In essence, this linearization method  can be viewed as a linear interpolation between two power flow solutions: the given operational point and a known operational point with no power injection. As a result, the resulted linear power flow model (\ref{lp1}) has better global approximation accuracy comparing with the standard linearized models based on local first-order Taylor expansion. As for the given operational point, depending on  practical conditions, it can be procured by 1) power flow calculation with sufficient grid information; 2) state estimation based on available measurements; 3) an feasible analytic power flow profile, e.g. the flat voltage solution with $\bv =\mb{1}\, p.u$. 
	In  \cite{linear_2}, a continuation analysis of the linear power flow model (\ref{lp1}) is performed on the IEEE 13-node system and a real feeder with about 2000 nodes, and the numerical results show that  the relative errors in voltages do not exceed 0.2\% and 0.6\%, respectively. Therefore, model (2) provides a good approximation of power flow for the proposed method to achieve efficient and accurate power aggregation.
}

\textit{Remark 1:} For exposition  simplicity, we outline the linear power flow model (\ref{lp1}) for a three-phase system. However, the proposed framework is clearly applicable to the systems with a mix of three-, double-, single-phase connections. For example, if a electric device is  double-phase or single-phase integrated, we fix the entries of the missing phases as zero in $\{\mb{p}_Y,\mb{q}_Y\}$ or $\{\mb{p}_\Delta,\mb{q}_\Delta\}$  and the corresponding line impedance matrix. Besides, 
the linear power flow model (\ref{lp1}) captures all possible connection manners of electric devices, and it is applicable to both meshed and radial distribution networks.

Accordingly, the network constraints can be formulated as 
\begin{subequations}  \label{net_con}
	\begin{align} 
	\underline{\mb{v}}\ &\leq \,\mb{v} \leq \bar{\mb{v}} \label{vol}\\
	\underline{\mb{i}}_L &\leq \mb{i}_{L} \leq \bar{\mb{i}}_L  \label{thermal}
	\end{align}
\end{subequations}
which involve the voltage limit constraints  (\ref{vol}) and the line thermal constraints (\ref{thermal}).


\subsection{Distributed Energy Resource Model}
We consider a discrete-time horizon $\mc{T}=\{1,2,\cdots,T\}$ and a variety of typical DERs, {including dispatchable PV units, energy storage  devices, directly controllable loads and HVAC systems. 
Based on references \cite{pv,pv2,esm0,esm00,esm1,esm2,hvac}, the DER operational models are built as follows.}

\subsubsection{Dispatchable PV Units}
$\forall i \in \mc{N}_{pv}, \ t\in \mc{T}$
\begin{subequations} \label{pv}
	\begin{gather} 
	\underline{P}_{i,t}^{g,\psi} \leq  P_{i,t}^{g,\psi} \leq \bar{P}_{i,t}^{g,\psi} \\
	\left(P_{i,t}^{g,\psi}\right)^2+\left(Q_{i,t}^{g,\psi}\right)^2 \leq \left(\bar{S}_{i,t}^{g,\psi}\right)^2
	\end{gather}
\end{subequations}
where $\mc{N}_{pv}$ denotes the set of buses connected with dispatchable PV units. 

\subsubsection{Energy Storage Devices}
$\forall i \in \mc{N}_{es}, \ t\in \mc{T}$
\begin{subequations} \label{bat}
   \begin{gather}
   \underline{P}_{i,t}^{e,\psi}  \leq  P_{i,t}^{e,\psi} \leq \bar{P}_{i,t}^{e,\psi} \\
   \left(P_{i,t}^{e,\psi}\right)^2+\left(Q_{i,t}^{e,\psi}\right)^2 \leq \left(\bar{S}_{i,t}^{e,\psi}\right)^2 \\
 {  E_{i,t}= \kappa_i \cdot E_{i,t-1}-\Delta t\cdot P_{i,t}^{e}} \label{dyn}\\
   \underline{E}_i  \leq E_{i,t} \leq \bar{E}_i   \label{soe}
\end{gather}
\end{subequations}
 Here,  $\mc{N}_{es}$ denotes the set of buses connected with energy storage devices and 
\begin{align*}
P_{i,t}^{e}:=\sum_{\psi}P_{i,t}^{e,\psi}
\end{align*}
is the total active power output (summation over the phases) of the ES devices at bus $i$ at time $t$, which can be either positive (discharging) or negative (charging).
 $\kappa_i\in(0,1]$ is the storage efficiency factor which models the energy loss over time, and $E_{i,0}$ denotes the  initial state of charge (SOC). { In (\ref{dyn}), {we assume  $100\%$} {charging and discharging} energy conversion efficiency for simplicity, i.e., no power loss in the charging or discharging process.
}


\subsubsection{Directly Controllable Loads }
$\forall i \in \mc{N}_{cl}, \ t\in \mc{T}$
\begin{subequations} \label{conload}
   \begin{gather}
   \underline{P}_{i,t}^{d,\psi} \leq  P_{i,t}^{d,\psi} \leq \bar{P}_{i,t}^{d,\psi}\\
   Q_{i,t}^{d,\psi}= \eta_{i}^{d}\cdot P_{i,t}^{d,\psi} \label{lo}
   \end{gather}
\end{subequations}
where $\mc{N}_{cl}$ denotes the set of buses connected with directly controllable loads. In (\ref{lo}), we assume fixed power factors with
constant $\eta_{i}^{d}$.


\subsubsection{HVAC Systems}
$\forall i\in \mc{N}_{hv}, \ t\in \mc{T}$
\begin{subequations} \label{hvac}
   \begin{gather}
      0 \leq  P_{i,t}^{h,\psi} \leq \bar{P}_{i,t}^{h,\psi} \\
   Q_{i,t}^{h,\psi}=\eta_{i}^{h}\cdot P_{i,t}^{h,\psi} \label{reac}\\
{ F^{in}_{i,t}=F^{in}_{i,t-1}+\alpha_i\cdot\left(F^{out}_{i,t}-F^{in}_{i,t-1}\right)+\frac{\Delta t}{\beta_i} \cdot P_{i,t}^{h}}\label{temp}\\
   \underline{F}_{i}\leq F^{in}_{i,t} \leq \bar{F}_{i} \label{comt}
\end{gather}
\end{subequations}
Here, $\mc{N}_{hv}$ denotes the set of buses connected with HVAC systems and 
$$P_{i,t}^{h}:=\sum_{\psi}P_{i,t}^{h,\psi}$$ is the total active HAVC load (summation over the phases) at bus $i$ at time $t$.
 In (\ref{reac}),  we assume fixed power factors with
constant $\eta_{i}^{h}$.
 Equation (\ref{temp}) depicts the indoor temperature dynamics,
 where $\alpha_i\in(0,1)$ and $\beta_i$ are the parameters specifying the thermal characteristics of the buildings and the environment. {{$\beta_i$ has the unit of heat capacity}}, and
 a positive (negative)  $\beta_i$ indicates that the HVAC appliances work in the heater (cooler) mode. We define $F^{in}_{i,0}$ as the initial indoor temperature.
 See \cite{hvac} for  detailed explanations. 
To facilitate the subsequent proof of disaggregation feasibility, we assume that the sign of $\beta_i$ keeps fixed during the considered time period, which implies that the HVAC appliances do not change their operation modes between cooling and heating.

{
\emph{Remark 2:} 
The DER models (\ref{pv})-(\ref{hvac}) are approximate for the trade-off between model precision and computational efficiency.
Take the energy storage model (\ref{bat}) for example. A more realistic model than (\ref{bat}) would have distinct charging and discharging efficiencies, which would render the model nonconvex and therefore hard to analyze and compute, due to the complementarity constraint for charging and discharging power \cite{esm2}.
 For computational efficiency, the simplified energy storage model (\ref{bat}) is generally acceptable in practice \cite{esm1,esm2,hvac}. Furthermore, the proposed power aggregation method is a generic framework applicable to the cases with various DER models that can be more detailed and realistic. }



\subsection{Incorporated Model}
{
{Accordingly, the vectors of three-phase active power injection via wye- and delta-connection at time $t$ are given by}
\begin{subequations}\label{dde}
	\begin{align}
	\mb{p}_{Y,t} \,:& =   \left\{ P_{i,t}^{g,\psi}, P_{i,t}^{e,\psi}, - P_{i,t}^{d,\psi}, -P_{i,t}^{h,\psi}     \right\}_{i\in \mathcal{N}_Y,\psi\in \phi_{Y}}\\
	\mb{p}_{\Delta,t} :& =   \left\{ P_{i,t}^{g,\psi}, P_{i,t}^{e,\psi}, - P_{i,t}^{d,\psi}, -P_{i,t}^{h,\psi}     \right\}_{i\in \mathcal{N}_\Delta,\psi\in \phi_{\Delta}}\\
	\mb{q}_{Y,t} \,:& =   \left\{ Q_{i,t}^{g,\psi}, Q_{i,t}^{e,\psi}, - Q_{i,t}^{d,\psi}, -Q_{i,t}^{h,\psi}     \right\}_{i\in \mathcal{N}_Y,\psi\in \phi_{Y}}  \\
		\mb{q}_{\Delta,t} :& =   \left\{ Q_{i,t}^{g,\psi}, Q_{i,t}^{e,\psi}, - Q_{i,t}^{d,\psi}, -Q_{i,t}^{h,\psi}     \right\}_{i\in \mathcal{N}_\Delta,\psi\in \phi_{\Delta}}  
	\end{align}
\end{subequations}
{Then we can construct the power injection vector $\mb{x}_t$, i.e., the realization of  $\mb{x}$  at time $t$, using definition} (\ref{dex}).}

As a result,
 the network model (\ref{lp1}) (\ref{net_con}) and the DER models (\ref{pv})-(\ref{hvac}) can be incorporated and equivalently rewritten as the following compact form:
\begin{subequations} \label{cf}
	\begin{align} 
\qquad\quad	P_{0,t}=  \mb{1}_3^\top\mb{D}\mb{x}_t&+\mb{1}_3^\top\mb{d}_t   & t\in\mc{T}    \label{p0} \\
   \sum_{\tau=1}^{t}\mb{H}_\tau \mb{x}_\tau&\leq \mb{h}_t  & t\in\mc{T} \label{tc} \\
   	\qquad\qquad\quad\mb{g}_t\left(\mb{x}_t,\mb{v}_t,\mb{i}_{L,t}\right)&\leq \mb{0}  & t\in\mc{T} \label{td}  
	\end{align}
\end{subequations}
Equation (\ref{p0}) is the formula of the net active power injection at the substation, which is obtained by summing (\ref{p_0}) over the three phases. 
 Equation (\ref{tc}) depicts the time-coupled constraints, involving the ES SOC limits (\ref{dyn}) (\ref{soe}) and the HAVC comfortable temperature limits (\ref{temp}) (\ref{comt}); matrix $\mb{H}_t$ and vector $\mb{h}_t$ are corresponding system parameters. 
Equation (\ref{td}) collects the time-decoupled operational constraints, and $\mb{g}_t\left(\cdot\right)$ is a vector-valued convex function. The power flow equalities (\ref{v}) (\ref{il}) are reformulated in an  equivalent unified form as inequalities, which are contained in (\ref{td}). { {
Note that $\mb{x}_t$ only contains the controllable power injection variables, i.e., those in} (\ref{dde}), {while the time-varying  uncontrollable loads and other non-dispatchable power generations are treated as given system parameters and captured by  $\mb{g}_t\left(\cdot\right)$ and $\mb{d}_t$. }
}


	\section{Power Aggregation Methodology}

In this section, we propose an inner-box approximation method to measure the aggregate flexibility. Then two optimization models are developed to implement power aggregation: one aims to evaluate the maximal flexibility level of the distribution systems, while the other is to optimally schedule  the flexibility reserve and the base-case power dispatch.

\subsection{Inner-box Approximation Method}

\begin{figure}[thpb]
	\centering
	\includegraphics[scale=0.26]{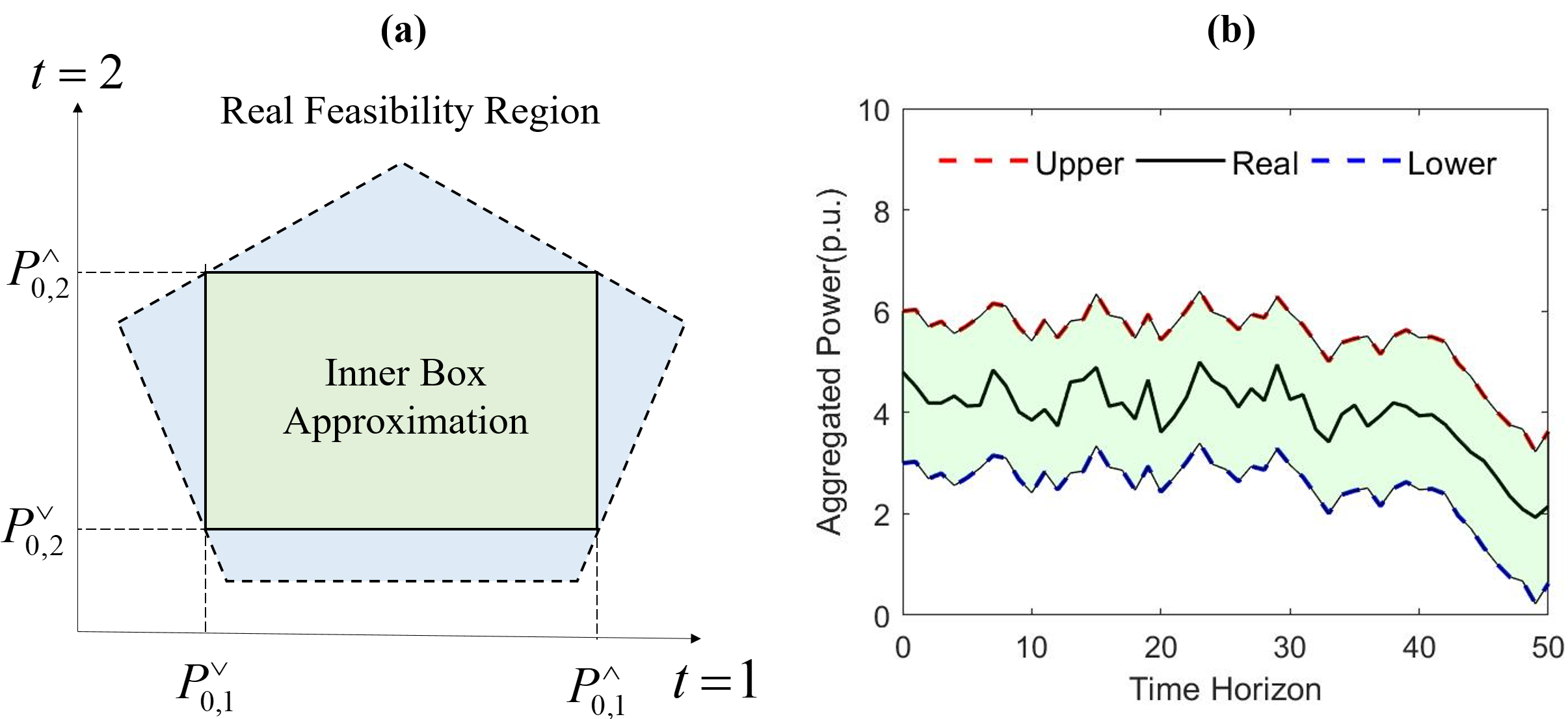}
	\caption{Inner-box approximation for the aggregated power feasible region. (a) illustrates the case with two time steps; (b) depicts how the actually deployed power trajectory is bounded by its upper and lower trajectories.}
	\label{def}
\end{figure}

Given the network and DER constraints, the goal of power aggregation is to determine the feasible region of the net power injection at the substation over time.
Since it is computationally impractical to procure the exact feasible region with massive DER devices, we quantify the aggregate flexibility  with an inner-box approximation method. As illustrated in Figure \ref{def}(a), we define a power interval $\left[{P}^\vee_{0,t}, {P}^\wedge_{0,t} \right]$ to restrict the net power injection for each time $t\in\mathcal{T}$, which forms a box-shape feasible region $\mathbb{S}$ in the power coordinates: 
\begin{align} \label{fr}
\begin{split}
\mathbb{S}&=\left[{P}^\vee_{0,1}, {P}^\wedge_{0,1}\right]\times\left[{P}^\vee_{0,2}, {P}^\wedge_{0,2}\right]\times\dots\times\left[{P}^\vee_{0,T}, {P}^\wedge_{0,T}\right] 
\end{split}
\end{align}

When mapped to the time coordinate, the feasible region $\mathbb{S}$ is specified by the upper power trajectory $\{ P^\wedge_{0,t}\}_{t\in\mathcal{T}} $ and the lower power trajectory $\{P_{0,t}^\vee\}_{t\in\mathcal{T}}$. As illustrated in Figure \ref{def}(b), the real trajectory of net power injection at the substation lies in the shade area given by the upper and lower ones. We further define the aggregate flexibility $E_{af}$ as 
\begin{align} \label{agg_f}
E_{af} := \sum_{t\in\mathcal{T}} \left( {P}^\wedge_{0,t}-{P}^\vee_{0,t}  \right)\cdot \Delta t
\end{align}
Notice that $E_{af}$ has the unit of energy, which is
 interpreted as the potential energy flexibility level of the distribution system. 

{
Comparing with the existing work, the proposed inner-box approximation method integrates three main advantages: 
\begin{itemize}

	\item [1)] (Compatible) Unlike aggregation approaches \cite{min_1,min_2,min_3} based on Minkowski sum of polytopic DER feasible sets, the proposed optimization-based method is compatible with various DER models and enables consideration of network constraints.
	\item [2)] (Safe) Comparing with  the outer approximation approach \cite{outer}, the proposed  method  provides a safe (conservative) approximation for the actual feasible region, 
 so  that an arbitrary trajectory within the inner-approximate feasible region $\mathbb{S}$ is achievable by properly coordinating the DER devices while respecting the operation constraints. 
 	\item [3)] (Simple) The feasible region $\mathbb{S}$ is defined in the form of time-decoupled intervals, which is simple and efficient for practical applications. In contrast to the transmission-distribution coordination methods \cite{td-1,td-2,td-3} using iterative decomposition algorithms, with the proposed method, the distribution grid can conveniently participate in transmission operation  by  reporting the concise and compact feasible region $\mathbb{S}$.
 
\end{itemize}
}

\textit{Remark 3:} Although the proposed method is illustrated for the case where the summation of net power injection across the phases is considered, it is capable of quantifying the power flexibility of each phase.

{In the following parts, the variables with 
 superscripts ``$\wedge$", ``$-$", ``$\vee$" correspond to
 the upper, base, and lower trajectories,  respectively. For example, 
${P}^\wedge_{0,t}, {P}^-_{0,t}, {P}^\vee_{0,t} $ are the net power injections at the substation for these three trajectories, respectively, which result from the corresponding nodal power injections ${\mb{x}}^\wedge_{t}, {\mb{x}}^-_{t}, {\mb{x}}^\vee_{t} $ by}
$$P_{0,t}^{\{\wedge,-,\vee\}}= \mb{1}_3^\top\mb{D}\mb{x}_t^{\{\wedge,-,\vee\}}+\mb{1}_3^\top\mb{d}_t.$$

\subsection{Maximal-flexibility Power Aggregation Model} 

With those definitions above in place, we aim to find the optimal approximate feasible region that achieves the largest aggregate flexibility. Accordingly, the following maximal-flexibility power aggregation (MPA) model (\ref{obj1})-(\ref{jc}) is built to solve the optimal upper and lower operational trajectories. 

\subsubsection{Objective Function}
\begin{equation} \label{obj1}
\max_{\mb{x}^\wedge_t,\mb{x}^\vee_t} \ E_{af} = \sum_{t=1}^T {\mb{1}_3}^\top\mb{D} \left( \mb{x}^\wedge_t -\mb{x}^\vee_t\right)\cdot\Delta t
\end{equation}
The objective (\ref{obj1}) is to maximize the aggregate flexibility $E_{af}$ (\ref{agg_f}) of the distribution system. 

\subsubsection{Individual Constraints }
\begin{subequations} \label{ic}
\begin{align}
\quad \sum_{\tau=1}^{t}\mathbf{H}_\tau\mathbf{x}_\tau^\wedge \leq \mathbf{h}_t,\ \  &\mb{g}_t\left(\mb{x}_t^\wedge,\mb{v}_t^\wedge,\mb{i}_{L,t}^\wedge\right)\leq \mathbf{0} \quad \forall t\in\mathcal{T} \label{ic_up}\\
\quad \sum_{\tau=1}^{t}\mathbf{H}_\tau\mathbf{x}_\tau^\vee \leq \mathbf{h}_t, \ \  &\mb{g}_t\left(\mb{x}_t^\vee ,\mb{v}_t^\vee ,\mb{i}_{L,t}^\vee \right)\leq \mathbf{0} \quad \forall t\in\mathcal{T} \label{ic_low}
\end{align} 
\end{subequations}
Identical to equation (\ref{tc}) (\ref{td}), equation (\ref{ic_up}) and (\ref{ic_low}) depict the network and DER constraints for the upper and lower operational trajectories respectively, where the inequalities are taken entry-wise. Since there is no overlapping term between the upper and lower trajectories in (\ref{ic_up}) and (\ref{ic_low}), we call them ``individual constraints".

\subsubsection{Joint Constraints}  
\begin{subequations} \label{jc}
\begin{align}
\qquad\qquad\qquad P_{0,t}^\vee\ & \leq P_{0,t}^\wedge \qquad\ \, \forall t\in\mathcal{T} \label{jc_ul} \\
\qquad P_{i,t}^{e,\wedge} &\leq P_{i,t}^{e,\vee}  \, \qquad \forall i\in\mathcal{N}_{es},t\in\mathcal{T}   \label{jc_bat}\\
\qquad P_{i,t}^{h,\vee} &\leq P_{i,t}^{h,\wedge} \qquad \forall i\in\mathcal{N}_{hv},t\in\mathcal{T} \label{jc_h}
\end{align}
\end{subequations}
{Equation (\ref{jc}) collects all the ``joint constraints" corresponding to the upper and lower trajectories, and it aims to guarantee the disaggregation feasibility, i.e., 
 any aggregate power trajectory between the upper and lower ones is achievable.}
In particular, equation (\ref{jc_ul}) indicates that the upper aggregate power should always be larger than the lower one, so that these two trajectories do not intersect with each other and the aggregate feasible region is well defined.
{
Equation (\ref{jc_bat}) implies that the ES power output associated with the upper trajectory should always be smaller than that at the lower trajectory. Intuitively, this equation is imposed to make any ES power trajectories within $[P_{i,t}^{e,\wedge}, P_{i,t}^{e,\vee}]_{t\in \mc{T}}$ satisfy the SOC limits (\ref{bat}d). Similarly,  equation (\ref{jc_h}) is utilized to guarantee that the comfortable temperature limits (\ref{comt}) will not be violated by any  HVAC load trajectories within $[P_{i,t}^{h,\vee}, P_{i,t}^{h,\wedge}]_{t\in \mc{T}}$. See Appendix B for the detailed explanations on how equation (\ref{jc})  contributes to the disaggregation feasibility.}

As a consequence, the proposed MPA model (\ref{obj1})-(\ref{jc}) 
is formulated as a quadratically constrained convex programming problem.
 Through solving this model, we can procure the largest inner-box approximation of the aggregate power feasible region, together with the optimal upper and lower operational trajectories. In addition, the disaggregation feasibility of any trajectories within this approximate region can be guaranteed, which is restated formally as the following proposition. The proof of Proposition \ref{pro1} is provided in Appendix B. 

\begin{proposition} \label{pro1}
Suppose that $\left\{P_{0,t}^{\vee}\right\}_{t\in\mathcal{T}}$  and $\left\{P_{0,t}^{\wedge}\right\}_{t\in\mathcal{T}}$ are the lower and upper aggregate power trajectories that respect the constraints (\ref{ic}) (\ref{jc}), then for any given trajectory $\left\{P_{0,t}^o\right\}_{t\in\mathcal{T}}$ that satisfies 
$P_{0,t}^{\vee} \leq  P_{0,t}^o \leq P_{0,t}^{\wedge} $ for all $ t\in\mathcal{T}$, there exists a disaggregation solution $\{\mathbf{x}_t^o\}_{t\in\mathcal{T}}$ satisfying the operational constraints (\ref{cf}) with $P_{0,t}^o= \mb{1}_3^\top\mb{D}\mb{x}_t^o+\mb{1}_3^\top\mb{d}_t$.
\end{proposition}


\textit{Remark 4:} Proposition 1 claims a generic property for any pair of lower and upper aggregate power trajectories that respect the constraints (\ref{ic}) (\ref{jc}), which is independent of the  objective function. Therefore, this property also works on the optimal lower and upper aggregate power trajectories $\left\{P_{0,t}^{\vee\star}, P_{0,t}^{\wedge\star}\right\}_{t\in\mathcal{T}}$ that solve the MPA model (\ref{obj1})-(\ref{jc}).

\subsection{Economic Power Aggregation Model}

Providing reserve services is one of the major schemes for distribution systems to offer flexibility support to the transmission grid. In the electricity market, the reward of flexibility reserve is based on the available power capacity instead of the actual regulated power. Accordingly, we define three operational trajectories (upper, base and lower) for power aggregation in the distribution system. The base trajectory is associated with the economic dispatch of the DER facilities, while the upper and lower trajectories are utilized to specify the reserves of upward and downward flexibility respectively. 

The following economic power aggregation (EPA) model (\ref{EPA})  is established to optimally schedule the power dispatch and flexibility reserve:
\begin{subequations} \label{EPA}
	\begin{align} 
	\text{Obj.}& \  \min_{\mathbf{x}_t^\wedge,\mathbf{x}_t^-,\mathbf{x}_t^\vee} \ \sum_{t=1}^T \left[ C_t\left( \mathbf{x}_t^- \right) - R_t\left( P_{0,t}^\vee, P_{0,t}^-, P_{0,t}^\wedge \right) \right]   \label{obj_e}\\
	\text{s.t.}&   \ \text{Equation  (\ref{ic}) (\ref{jc_bat}) (\ref{jc_h})}   \label{dup}\\
	& \sum_{\tau=1}^{t}\mathbf{H}_\tau\mathbf{x}_\tau^- \leq \mathbf{h}_t,\  \mb{g}_t(\mb{x}_t^- ,\mb{v}_t^- ,\mb{i}_{L,t}^- )\leq \mathbf{0} \ \ \ \forall t\in\mathcal{T} \label{base_e}\\
	&  P_{0,t}^\vee \leq P_{0,t}^- \leq P_{0,t}^\wedge\qquad\qquad\qquad\qquad \quad  \  \ \forall t\in\mathcal{T} \label{base_p}
	\end{align}
\end{subequations}
In objective function (\ref{obj_e}), $C_t$ is the cost function associated with the base trajectory, and $R_t$ is the reward function for the flexibility reserve. The detailed formulation of  $C_t$ is
\begin{align}
\begin{split}
C_t&:=  \sum_{i\in\mathcal{N}_{es}} c_{i}^e\cdot(P_{i,t}^{e,-})^2 +\sum_{i\in\mathcal{N}_{hv}} c_i^{h}\cdot(T_{i,t}^{in,-} - T_i^{cf})^2 \\
&+\sum_{i\in\mathcal{N}_{pv}}\left[ c_{1,i}^{pv}\cdot{P_{i,t}^{g,-}}+ c_{2,i}^{pv}\cdot({P_{i,t}^{g,-}-\bar{P}_{i,t}^{g}})^2  \right] +p_t\cdot P_{0,t}^-  \label{cost}
\end{split}
\end{align}
where the first term captures the damaging effect of  charging/discharging to the ES systems. The second term describes the HVAC disutility of deviating from the most comfortable temperature $T_i^{cf}$. The third term denotes the operational cost and the power curtailment cost of PV units. $c_i^e, c_i^h, c_{1,i}^{pv}, c_{2,i}^{pv}$ are the corresponding cost coefficients. The last term is the cost of purchasing electricity from the transmission grid with the real-time price $p_t$. As for the  reward function $R_t$, the concrete definition is 
\begin{align} \label{reward}
R_t: =  r_t^\wedge \cdot(P_{0,t}^\wedge-P_{0,t}^-) + r_t^\vee\cdot(P_{0,t}^--  P_{0,t}^\vee)
\end{align}
where $r_t^\wedge$ and $r_t^\vee$ are the reward coefficients for upward and downward flexibility reserve at time $t$ respectively.

Comparing with the MPA model, the EPA model aims to minimize the net operation cost of the distribution system, and we supplement constraints (\ref{base_e}) and (\ref{base_p}) to restrict the base operational trajectory in a similar way. 
In (\ref{dup}), 
equation (\ref{ic}) (\ref{jc_bat}) (\ref{jc_h}) are duplicated from the MPA model. 
 By solving the EPA model (\ref{EPA}), distribution system operators can obtain the economically optimal power dispatch schemes $\left\{ \mathbf{x}_t^{-\star} \right\}_{t\in\mathcal{T}}$ and flexibility reserve intervals $[ P_{0,t}^{\vee\star},P_{0,t}^{\wedge\star}]_{t\in\mathcal{T}}$ simultaneously. Meanwhile, the distribution system provides reserve services for the transmission grid by reporting its flexibility intervals. According to Proposition 1, the distribution system is capable of tracking any  power regulation signals from the transmission side that are within the flexibility intervals .

{\emph{Remark 5:} 
Note that the base DER power trajectories are not restrained to lie between their upper and lower ones in  model (\ref{EPA}). Indeed, constraints (\ref{jc_bat}) (\ref{jc_h}) and (\ref{base_p}) are sufficient to guarantee the disaggregation feasibility (see Proposition 1). The simulation results in Section \ref{si-EPA} further verify this point.
}

	\section{Distributed MPC Framework for Transmission-Distribution Interaction} \label{sec:dec MPC}

This section presents the transmission-distribution interaction framework with the proposed power aggregation method. To protect private information of participating DER facilities and enable scalable application, we develop a distributed MPC solver for the practical implementation of this framework.
 
\subsection{ Transmission-Distribution MPC Interaction Framework}                                                
{\color{black} 
In essence, power aggregation can be regarded as a trip planner for the future operational trajectories of the distribution system. 
Since the operations of energy storage devices and HVAC facilities are highly time-coupled, the optimally planned  trajectories may be significantly changed by executing the upcoming regulation commands from the transmission, so that a new round of power aggregation is required to update the scheduling after each transmission-distribution interaction. 
In addition, it is arduous to accurately forecast the renewable generation and uncontrollable power demand  for a long term. Hence, we implement the transmission-distribution interaction in the MPC manner \cite{mpc}. 
In this way, future time slots are taken into consideration when making the current decisions, and latest updated information can be utilized in each interaction.}
\begin{figure}[thpb]
	\centering
	\includegraphics[scale=0.322]{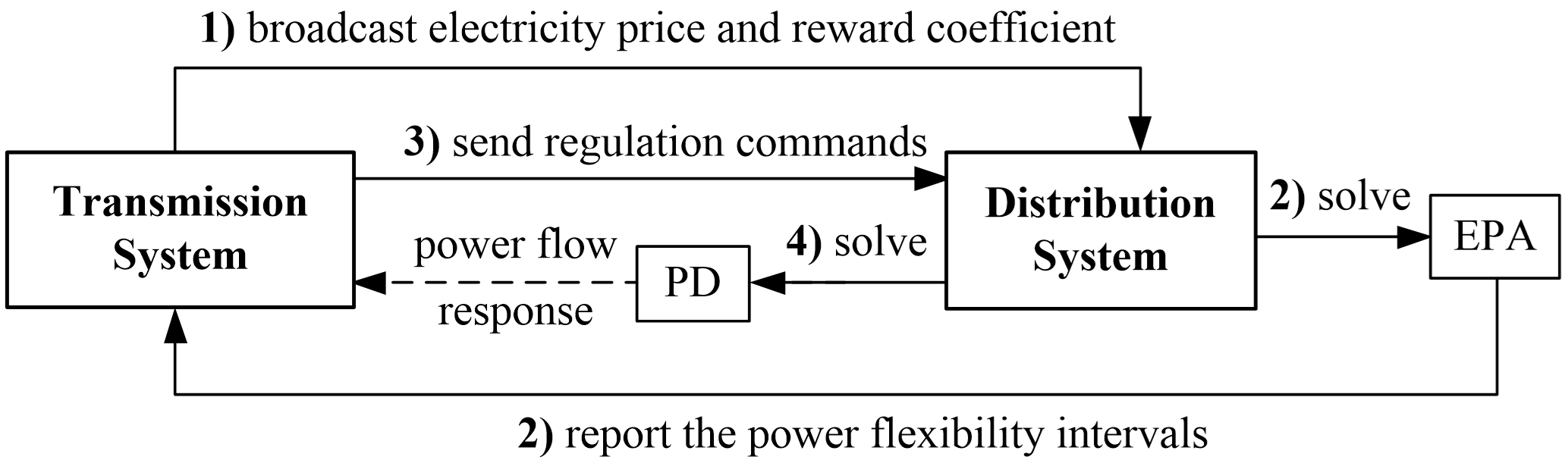}
	\caption{Schematic of the transmission-distribution interaction framework.}
	\label{frame}
\end{figure}

The framework for transmission-distribution interaction is proposed as the following procedure, which is illustrated as Figure \ref{frame}.

\begin{itemize}
	\item [1)] The transmission broadcasts the time-varying electricity price $p_t$ and reward coefficients $r_t^\wedge,r_t^\vee $ for the next $T_p$ time steps to each connected distribution system. 
	\item [2)] Based on the broadcast information, each distribution system performs the EPA model (\ref{EPA}) w.r.t. the next $T_p$ time steps, and report its solved flexibility interval $\left[P_{0,t}^{\vee,\star}, P_{0,t}^{\wedge,\star} \right] $ to the transmission.
	\item[3)] After gathering the flexibility intervals and the generators' information, the transmission determines the optimal power dispatch schemes and the regulation commands $P_{0,t}^{reg}\in \left[P_{0,t}^{\vee,\star}, P_{0,t}^{\wedge,\star} \right]$ in the next $T_d$ time steps ($T_d$ is typically much smaller than $T_p$) for each distribution system.  
	
	\item[4)] Once receiving the regulation command, each distribution system solves the following power disaggregation (PD) problem (\ref{PD}), and executes the decisions to optimally track the regulation command and dispatch its DER facilities for the next $T_d$ time steps.
	\begin{subequations} \label{PD}
		\begin{align} 
		\text{Obj.}& \ \ \min_{\mathbf{x}_t} \ \sum_{t=1}^{T_d} \left[ C_t\left( \mathbf{x}_t \right) + \rho_t\cdot(P_{0,t}-P_{0,t}^{reg} )^2    \right]
		  \label{pd:obj_e}\\
		\text{s.t.}&  \text{ Equation (\ref{cf})}
		\end{align}
	\end{subequations}
	where $\rho_t$ is the penalty coefficient for deviating from the regulation command.
	\item [5)]  {\color{black} Move $T_d$ time steps forward and 
	  repeat from step 1) with the updated DER conditions.}
\end{itemize}

\subsection{ Distributed Solution Algorithm }

To perform the above framework, a centralized solver needs to gather the real-time operational information of all DER facilities and solve a global optimization problem. Not only does it carry  huge computational and communication burdens, but also violate the privacy of each participating DER stakeholder. { To address these issues, we develop a distributed solver for the MPA, EPA and PD models based on {the predictor corrector proximal multiplier (PCPM)}  algorithm \cite{pcpm}.}

Let $\mb{y}_0$ and $\{\mb{y}_k\}_{k\in\mathcal{K}}$ denote the local variables of the system aggregator (SA) and $k$-th DER facility for the considered period $\mc{T}_c$, which are defined as follows.
\begin{subequations}
	\begin{align}
	& \mb{y}_0=\left\{ \mb{v}_t^{\wedge,-,\vee},\,\mb{i}_{L,t}^{\wedge,-, \vee},\,P_{0,t}^{\wedge,-,\vee} \right\}_{t\in\mc{T}_c}\in \mc{Y}_0 &\\
	& \mb{y}_k=\left\{\mb{x}_{k,t}^{\wedge,-,\vee} \right\}_{t\in\mc{T}_c} \in \mc{Y}_k \qquad \quad \  \forall k\in\mc{K}
	\end{align}
\end{subequations}
where $\mc{K}:=\{1,2,\cdots,K\}$ is the index set of DER facilities. $\mc{Y}_0$ and $\mc{Y}_k$ are the feasible sets for the corresponding local variables, which are specified by their own operational constraints. 
Leveraging their separable structure, the MPA, EPA and PD models can be rewritten as the compact form (\ref{comp}):
\begin{subequations}
	\begin{align}
	\text{Obj.}  & \   \min_{\mb{y}_0\in\mc{Y}_0,\mb{y}_k\in\mc{Y}_k} 
	f_0(\mb{y}_0 )+\sum_{k=1}^K f_k(\mb{y}_k)  \label{obj_3} \\
	\text{s.t.} &\quad  \mb{y}_0 = \sum_{k=1}^{K}\mb{W}_k\mb{y}_k + \mb{w} \label{com_3}
	\end{align} \label{comp}
\end{subequations}
where $f_0$ and $f_k$ captures the corresponding objective functions for the SA and $k$-th DER facility.
 Constraint (\ref{com_3}) represents the power flow equations (\ref{lp1}), while matrix $\mb{W}_k$ and vector $\mb{w}$ are system parameters.

\begin{figure}[thpb]
	\centering
	\includegraphics[scale=0.295]{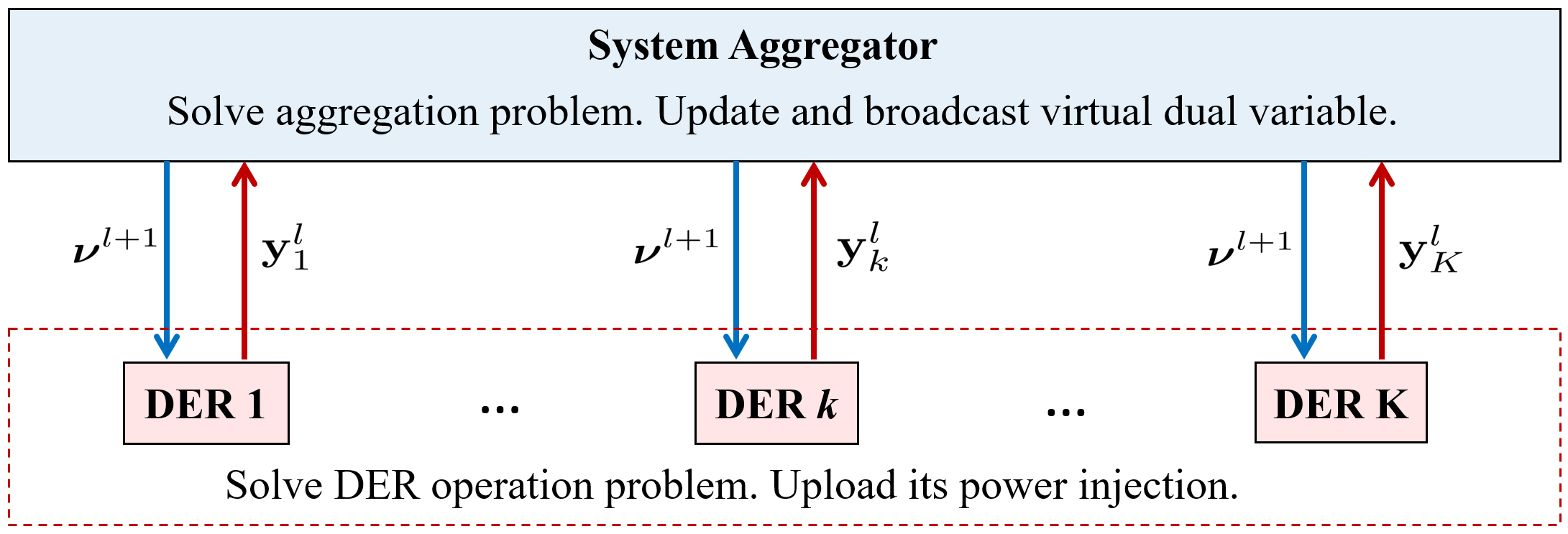}
	\caption{Schematic of the distributed solution algorithm.}
	\label{PCPM}
\end{figure}

Introducing the dual variable $\bm{\mu}$ and virtual dual variable $\bm{\nu}$, we develop the distributed solution algorithm as Algorithm 1. The implementation procedure is illustrated as Figure \ref{PCPM}. In each iteration, it requires two-way communications between the SA and every DER facility, and each  entity individually solves its  small-scale optimization problem in parallel.

\begin{algorithm}[H]
	\caption{: Distributed Solution Algorithm}
	\textbf{1. Initialization:} 
	$l\leftarrow 0$. Each DER facility $k\in\mc{K}$ sets its initial $\mb{y}_k^{0}$ and sends to the system aggregator (SA). The SA sets the initial $\mb{y}_0^{0}$, $\bm{\mu}^{0}$, step length $\rho$ and tolerance $\epsilon$.
	
	\textbf{2. Update Virtual Dual Variables:}    
	The SA updates the virtual dual variables by
	\begin{align}
	\bm{\nu}^{l+1}=\bm{\mu}^l + \rho \left( \sum_{k=1}^K\mb{W}_k\mb{y}_k^{l} + \mb{w} -\mb{y}_0^{l} \right)
	\end{align}
	then broadcast $\bm{\nu}^{l+1}$ to every DER.
	
	\textbf{3. Parallel Optimization:}
	
	For SA: solve the power aggregation problem
	\begin{align}\label{sai}
	\mb{y}_0^{l+1}=\arg \min_{\mb{y}_0\in\mc{Y}_0} 
	f_0(\mb{y}_0 ) - {\bm{\nu}^{l+1}}^\top\mb{y}_0
	+ \frac{1}{2\rho}||\mb{y}_0-\mb{y}_0^{l} ||^2 
	\end{align}
	
	For DER $k$: solve its DER operation problem
	\begin{align}\label{deri}
	\mb{y}_k^{l+1}=\arg \min_{\mb{y}_k\in\mc{Y}_k}  f_k(\mb{y}_k) +
	{\bm{\nu}^{l+1}}^\top\mb{W}_k\mb{y}_k 
	+ \frac{1}{2\rho}||\mb{y}_k-\mb{y}_k^{l} ||^2 
	\end{align}
	and send $\mb{y}_k^{l+1}$ to SA. 
	
	\textbf{4. Update Dual Variables:} The SA updates the dual variables by
	\begin{align}
	\bm{\mu}^{l+1}=\bm{\mu}^l + \rho \left( \sum_{k=1}^K\mb{W}_k\mb{y}_k^{l+1} + \mb{w} -\mb{y}_0^{l+1} \right)
	\end{align}
	
	\textbf{5. Check Convergence:} if $||\bm{\mu}^{l+1}-\bm{\mu}^l||\leq \epsilon$, terminate. Otherwise $l\leftarrow l+1$ and go back to step \textbf{2}.
\end{algorithm}

{ 
{
As for the convergence, it has been shown in} \cite{pcpm} that 
as long as strong duality holds for the original optimization problem,
the distributed algorithm will converge to the optimal solution when  the step length $\rho$ is  positive and smaller than a  threshold.
Since MPA, EPA and PD are all convex optimization problems which are assumed to be strictly feasible (indicating strong duality), convergence of the distributed algorithm is guaranteed with a sufficiently small step length.
Numerical results in Section \ref{nu-con} further validate that the distributed algorithm  converges fast for a real-world distribution system.
}

	
\section{Numerical Tests}

\subsection{Simulation Setup}

\begin{figure}[thpb]
	\centering
	\includegraphics[scale=0.32]{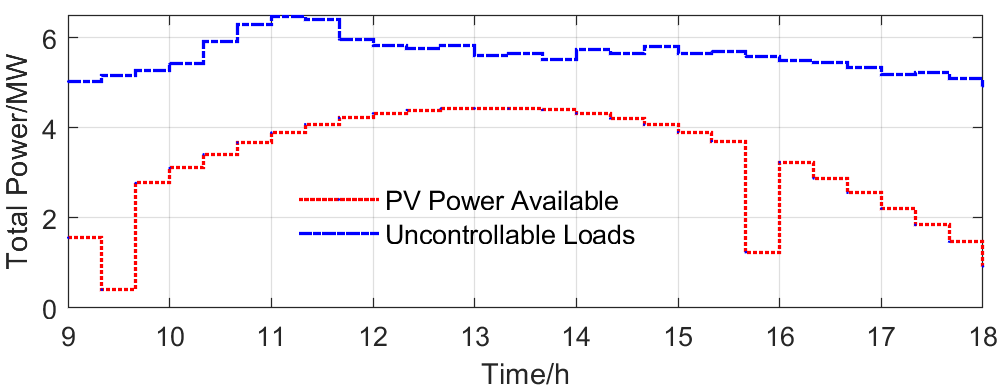}
	\caption{Total PV power available and uncontrollable loads from 9:00 to 18:00 with the granularity of 20 minutes.}
	\label{L_PV}
\end{figure}

Numerical tests are carried out on a real distribution feeder located within the territory of  Southern California Edison (SCE). This feeder contains 126 multi-phase buses with a total of 366 single-phase connections. The nominal voltage at the substation is 12kV (1 p.u.), and we set the upper and lower limits of voltage magnitude as 1.02 p.u. and 0.98 p.u. Dispatchable DERs include 33 PV units, 28 energy storage devices and 5 HVAC systems. The real data of power consumption from industrial, commercial, and residential loads are applied, as well as real solar irradiance profiles. The total amounts of uncontrollable loads and PV available power from 9:00 to 18:00 are presented as Figure \ref{L_PV}. {For PV units, we set   the lower bound of power generation as zero and take the PV available power in Figure \ref{L_PV} as the upper bound.}
The simulation time is discretized with the granularity of 20 minutes. We set the initial SOC of the energy storage devices to $50\%$. Detailed configurations and parameters of this feeder system are provided in \cite{sys}.

\subsection{Implementation of Maximal-flexibility Power Aggregation}

\begin{figure}[thpb]
	\centering
	\includegraphics[scale=0.32]{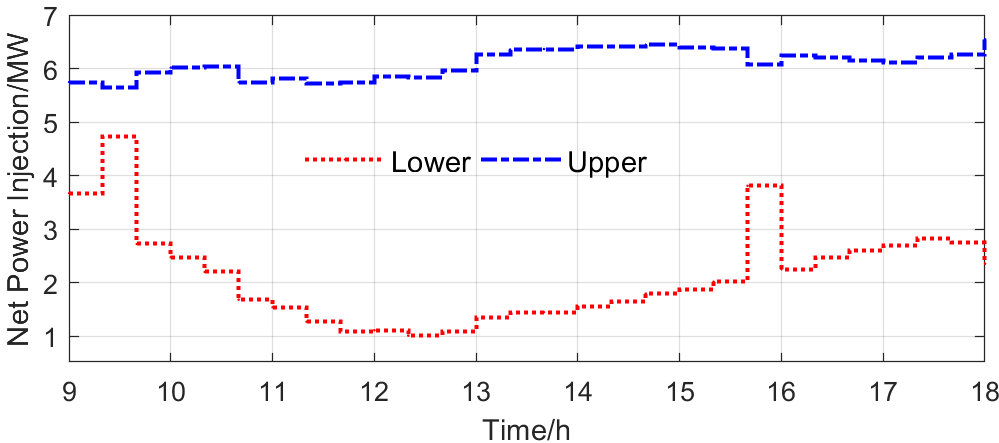}
	\caption{Upper and lower trajectories of net power injection at substation via MPA model.}
	\label{P0}
\end{figure}

\begin{figure}[thpb]
	\centering
	\includegraphics[scale=0.32]{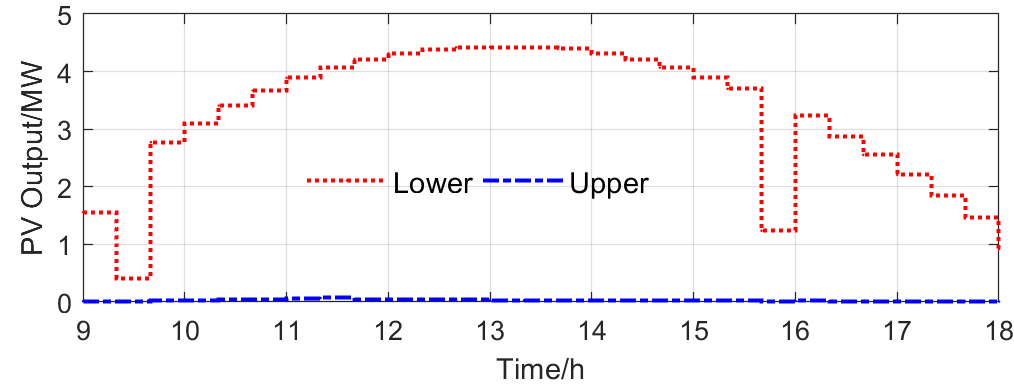}
	\includegraphics[scale=0.32]{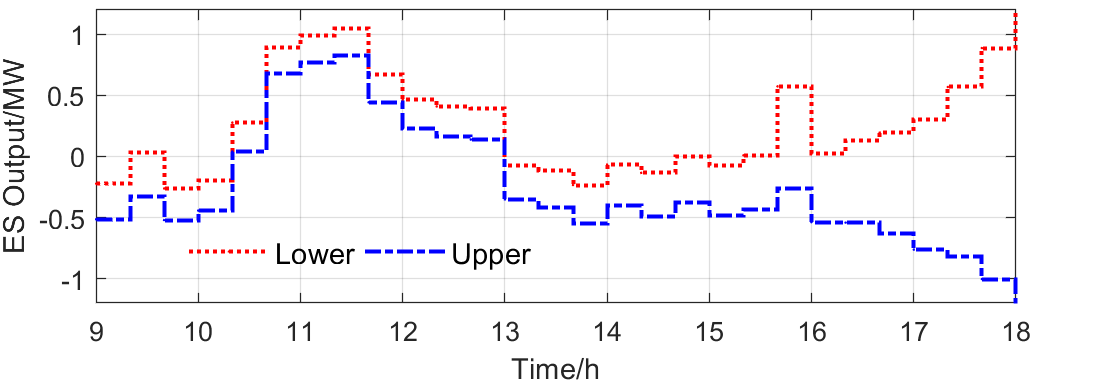}
	\includegraphics[scale=0.32]{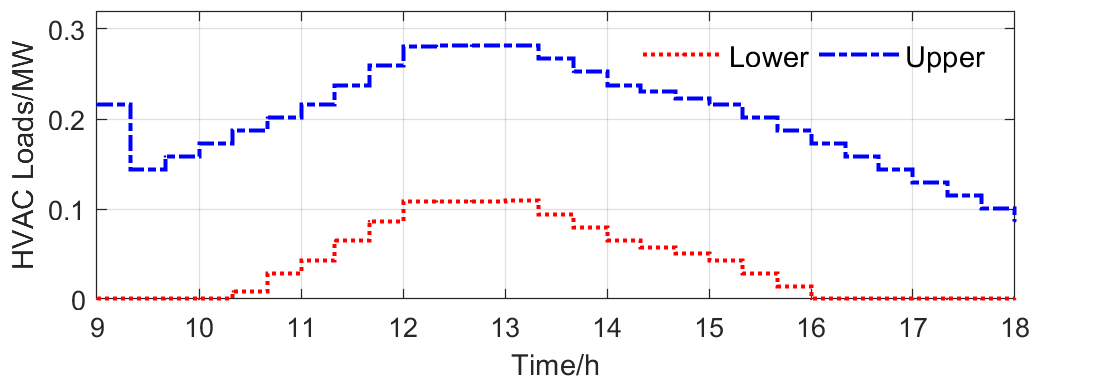}
	\caption{Upper and lower operation trajectories of DER facilities via MPA model.}
	\label{PV}
\end{figure}

We implemented the MPA model (\ref{obj1})-(\ref{jc}) to evaluate the maximal flexibility of the test system. The optimal upper and lower trajectories of the net power injection at the substation are presented as Figure \ref{P0}. The aggregate flexibility, i.e., the area between the upper and lower trajectories, is calculated as $E_{af}=37.1$ MW$\cdot$h. The corresponding DER power dispatch schemes are shown as Figure \ref{PV}. It is observed that the upper trajectory corresponds to less power injection and larger amount of loads comparing with the lower one, which is consistent with the intuition.


{In addition, Monte Carlo simulations are performed to verify the disaggregation feasibility of any  power trajectories within the intervals $\left[P_{0,t}^{\vee,\star}, P_{0,t}^{\wedge,\star} \right]_{t\in \mathcal{T}} $ in Fig. \ref{P0}. We assume that the regulated power trajectory $\{ P_{0,t}^{reg}\}_{t\in \mathcal{T}}$ from the transmission are random variables following the uniform  distribution, i.e., $P_{0,t}^{reg} \sim \text{Unif}\left(P_{0,t}^{\vee,\star}, P_{0,t}^{\wedge,\star} \right)$ independently for each $t\in \mathcal{T}$. 
Up to 5000 regulated power trajectories are  randomly generated, and we solve the following  disaggregation feasibility problem with $P_{0,t}=P_{0,t}^{reg}$ for each case.
\begin{align} \label{fess}
    \min \,  0\quad 
    s.t. \  \text{Equation} (\ref{cf})
\end{align}
The simulation results show that the problem (\ref{fess}) is feasible for all the generated power trajectories $\{ P_{0,t}^{reg}\}_{t\in \mathcal{T}}$ and 
 we can always obtain a corresponding disaggregation solution $\{\bx_t\}_{t\in \mathcal{T}}$ for each of them. These results are consistent with Proposition 1 where we theoretically show the disaggregation feasibility.

}

\subsection{Implementation of Transmission-Distribution Interaction} \label{si-EPA}

We further carry out the transmission-distribution interaction framework for the test system, which is implemented with the proposed MPC setting. At each time instant, the EPA model is performed for the next 12 time steps (4 hours), while only the flexibility intervals of the first time step are reported, i.e., $T_p=12$ and $T_d =1$, before the procedure moves one time step forward. The regulation commands $P_{0,t}^{reg}$ sent by the transmission are generated randomly and independently across time $t$ around the optimal operating points $P_{0,t}^{-,\star}$ with Gaussian distribution, and are projected onto $\left[P_{0,t}^{\vee,\star}, P_{0,t}^{\wedge,\star} \right]$. We set the penalty coefficient $\rho_t$ as a very large value for simplification, so that the test feeder system strictly tracks the regulation commands.

\begin{figure}[thpb]
	\centering
	\includegraphics[scale=0.315]{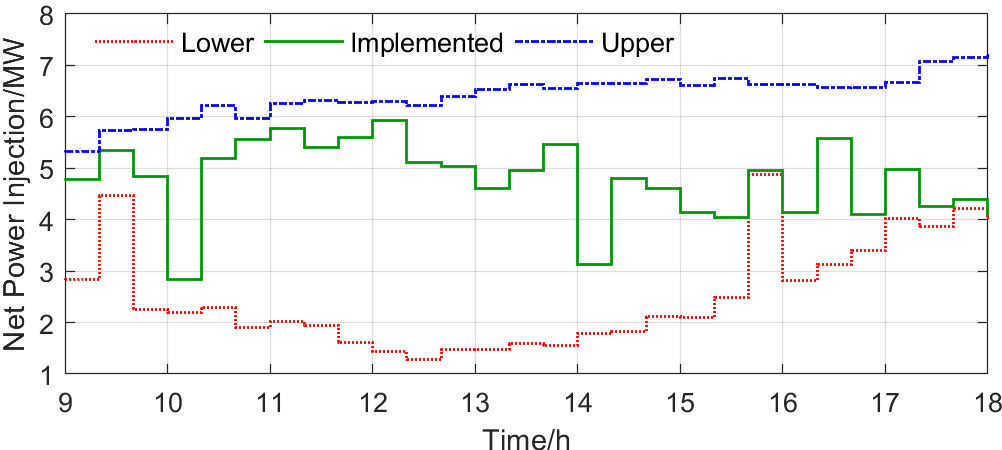}
	\caption{Upper, implemented, lower trajectories of net power injection at substation via transmission-distribution interaction framework.}
	\label{P0_E}
\end{figure}

\begin{figure}[thpb]
	\centering
	\includegraphics[scale=0.315]{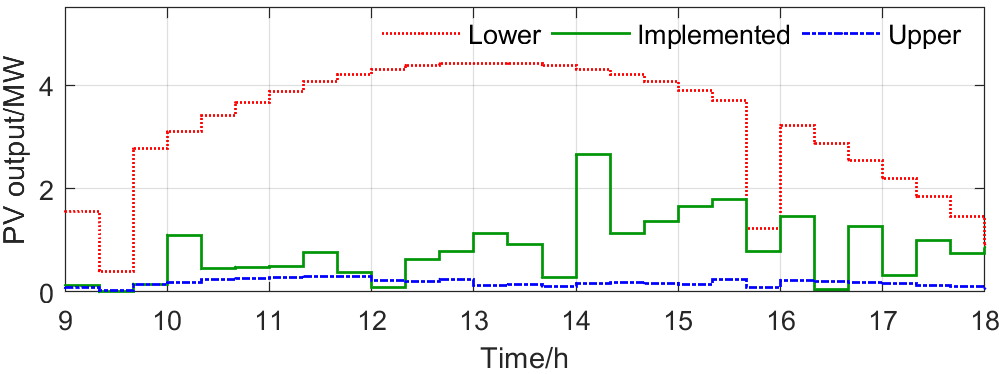}
	\includegraphics[scale=0.315]{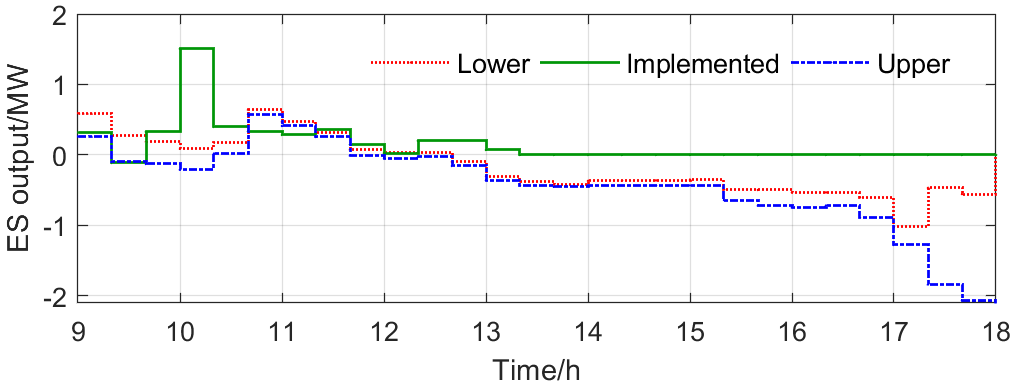}
	\includegraphics[scale=0.315]{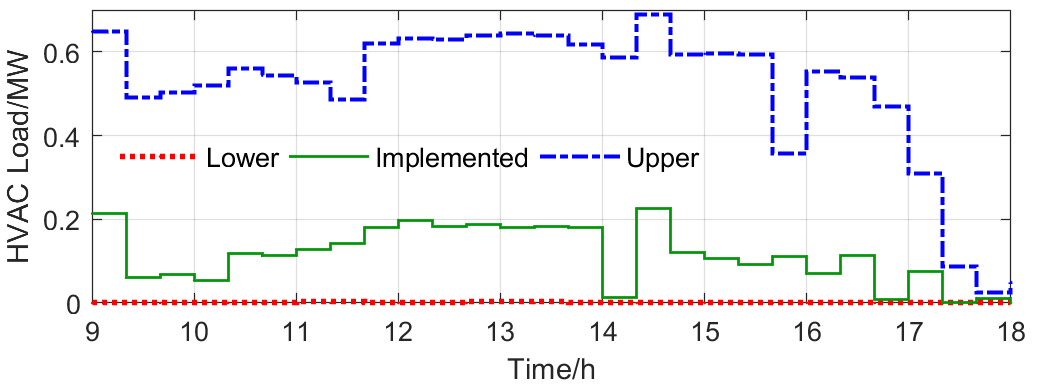}
	\caption{Upper, implemented, lower power trajectories of DER facilities via transmission-distribution interaction framework. }
	\label{PV_E}
\end{figure}

Figure \ref{P0_E} illustrates the upper, implemented, lower trajectories of the net power injection at the substation. The dashed red and blue curves represent the reported flexibility intervals  $\left[P_{0,t}^{\vee,\star}, P_{0,t}^{\wedge,\star} \right]$, and the solid green curve denotes the eventually implemented power trajectory, i.e., the regulation commands $P_{0,t}^{reg}$. These trajectories are accumulated step by step in the receding horizon. Figure \ref{PV_E} presents the DER power dispatch schemes associated with the three trajectories. It is observed that the implemented DER trajectories do not always lie between their upper and lower trajectories, while it is still ensured that the implemented aggregate power is within the reported flexibility intervals.

\subsection{Convergence of Distributed Solver} \label{nu-con}

\begin{figure}[thpb]
	\centering
	\includegraphics[scale=0.33]{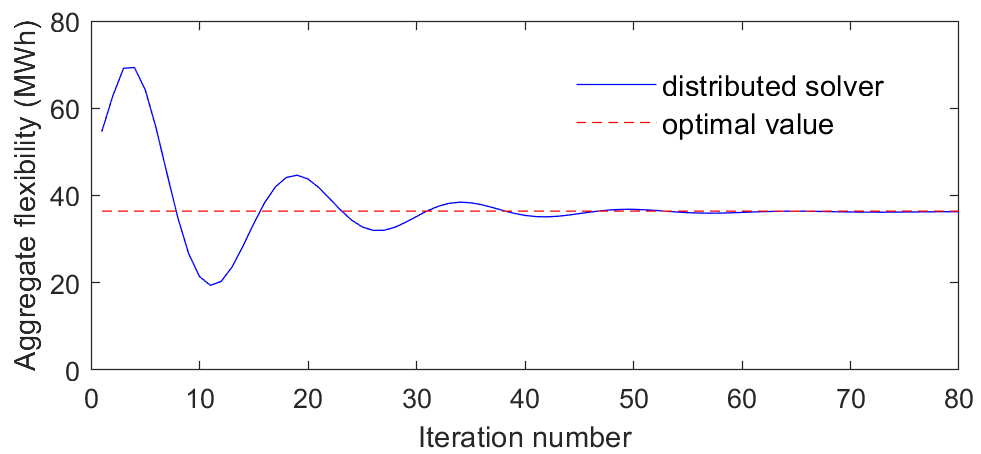}
	\caption{Convergence of the distributed solver for the MPA model.}
	\label{conver}
\end{figure}

In the above simulations, the distributed solver developed in Section \ref{sec:dec MPC}-B is used to solve the MPA, EPA and PD models. Taking the MPA model as an example, the convergence of the distributed solver is shown in Figure \ref{conver}. It is observed that this solver converges within tens of iterations and solves the MPA model efficiently.

\subsection{Computational Efficiency}
{ 

{Numerical simulations} are performed in a computing environment with Intel(R) Core(TM) i7-7660U CPUs running at 2.50 GHz and with 8-GB RAM. 
All the programmings are implemented in Matlab 2017b, while we use the CVX package \cite{cvx1} to model the convex programs and solve them with Gurobi optimizer 6.0 \cite{gur}. A distributed solver based on Algorithm 1 and a centralized solver are developed to solve the MPA and EPA model. 

For the 126-bus test system with 27 time periods, the computation time of the centralized solver and the distributed solver  is presented in Table \ref{ct}. For the distributed solver, ``successively" and ``in parallel" indicate whether
the individual optimization problems (\ref{sai}) (\ref{deri}) in step 3 of Algorithm 1 are solved one by one or in parallel. 
From Table \ref{ct}, it is observed that the distributed solver takes longer time than the centralized solver, because the distributed solver performs an iterative process to obtain the optimal solution. Besides, the computation time of each iteration is  dominated by the solution of the SA problem (\ref{sai}). For example, when solving the MPA model, it takes 16.76s on average for the distributed solver to perform one iteration ``successively", in which 9.55s is spent on solving the SA problem. That is because SA is responsible for 
solving the network optimization problem and thus deals with a much larger problem than each DER. In practice, SA is usually equipped with high  computational power, 
which enables a fast distributed solution process with parallel optimization.

\begin{table}[!t]
\renewcommand{\arraystretch}{1.3}
\caption{Computation time for  centralized solver and  distributed solver to solve  MPA and EPA model.}
\label{ct}
\centering
\begin{tabular}{cccc}
\hline\hline
\multirow{2}{*}{} & \multirow{2}{*}{$\ \ $ Centralized $\ \ $} & \multicolumn{2}{c}{$\ \ $Distributed $\ \ $} \\ \cline{3-4} 
    &     & $\ $successively $\ $   & $\ $ in parallel $\ $   \\ \hline
$\ $ MPA $\ $& 87.5s & 754.2s &  429.8s \\ 
EPA & 146.6s   &  1314.8s  &    661.2s  \\ \hline \hline
\end{tabular}
\end{table}

}

\section{Conclusion}

This paper develops a method to model and quantify the aggregate flexibility for different types of DERs in unbalanced distribution systems. With the inner-box approximation approach, we use the upper and lower aggregate power trajectories to specify the feasible region of the net power injection at the substation. Two convex optimization models are established to implement distribution-level power aggregation, which incorporate the network constraints and multi-phase unbalanced modeling. In addition, a  distributed MPC framework is proposed to implement the transmission-distribution interaction in a scalable and privacy-preserving manner. The effectiveness of our proposed method are validated via the numerical tests on a real distribution feeder.

\appendices
\section{Linear Multi-Phase Power Flow Model}
{ 
{Define $\mb{x}_Y:=\left[\mb{p}_Y^\top,\mb{q}_Y^\top \right]^\top$ and $\mb{x}_\Delta:=\left[\mb{p}_\Delta^\top,\mb{q}_\Delta^\top \right]^\top$},  thus we have $\mb{x}:=\left[{\mb{x}_Y}^\top, {\mb{x}_\Delta}^\top \right]^\top$. Let $\tilde{\bv}$ be the column vector collecting the three-phase nodal complex voltage.
Based on the given operational point $\{\tilde{\mb{v}}^o,{\mb{x}}_Y^o,{\mb{x}}_\Delta^o \}$, we can derive the following linear formulation (\ref{com_v}) for $\tilde{\bv}$ with the fixed-point linearization method \cite{linear_1}:
\begin{align} \label{com_v}
\tilde{\bv}= \mb{M}_Y\mb{x}_Y + \mb{M}_\Delta\mb{x}_\Delta + \mb{m}
\end{align}
where $\mb{M}_Y$, $\mb{M}_\Delta$ and $\mb{m}$ are defined as 
\begin{align*}
\mb{M}_Y &:= \left[\mathbf{Y}_{LL}^{-1}\text{diag}({\tilde{\mathbf{v}}}^{o*})^{-1}, -\jmath\cdot\mathbf{Y}_{LL}^{-1}\text{diag}({\tilde{\mathbf{v}}}^{o*})^{-1} \right] \\
\mb{M}_\Delta &:= \left[ \mathbf{Y}_{LL}^{-1}\mathbf{L}^\top \text{diag}(\mathbf{L}{\tilde{\mathbf{v}}}^{o*})^{-1},-\jmath\cdot \mathbf{Y}_{LL}^{-1}\mathbf{L}^\top \text{diag}(\mathbf{L}{\tilde{\mathbf{v}}}^{o*})^{-1} \right] \\
\mb{m} &:= -\mathbf{Y}_{LL}^{-1}\mathbf{Y}_{L0}\tilde{\mathbf{v}}_0
\end{align*}
Here, $\tilde{\mathbf{v}}_0\in\C^3$ denotes the three-phase complex voltage at the substation bus. $\mb{Y}_{LL}\in \C^{3N\times3N}$ is the sub-matrix of the three-phase admittance matrix
\begin{equation*}
\mathbf{Y}:= \begin{bmatrix} \mathbf{Y}_{00} & \mathbf{Y}_{0L} \\ \mathbf{Y}_{L0} &\mathbf{Y}_{LL} \end{bmatrix} \in \C^{3(N+1)\times3(N+1)}
\end{equation*}
$\mathbf{L}$ is a block-diagonal matrix defined by
\begin{equation*}
\mathbf{L}:= \begin{bmatrix} \mathbf{T} & & \\ & \ddots & \\ & & \mathbf{T} \end{bmatrix}, \quad 
\mathbf{T}:= \begin{bmatrix} 1 & -1  & 0 \\ 0 & 1 & -1 \\ -1 & 0 & 1 \end{bmatrix}
\end{equation*}

To derive the linear model for the voltage magnitudes $\bv = |\tilde{\bv}|$, we leverage the following derivation rule
\begin{align*}
\frac{\partial |f(x)|}{\partial x} = \frac{1}{|f(x)|} \mathcal{R} \left\{ f(x)^* \frac{\partial f(x)}{\partial x}  \right\}
\end{align*}
and obtain equation (\ref{v}),
where $\mb{A}:=\left[\mb{A}_Y,\mb{A}_\Delta \right]$ with 
\begin{align*}
\mb{A}_Y&:= \frac{\partial |\tilde{\bv}|}{\partial \bx_Y}= \text{diag}(|{\tilde{\mathbf{v}}}^{o}|)^{-1}\mathcal{R}\left\{ \text{diag}({\tilde{\mathbf{v}}}^{o*})\mb{M}_Y  \right\}\\
\mb{A}_\Delta&:=\frac{\partial |\tilde{\bv}|}{\partial \bx_\Delta}=\text{diag}(|{\tilde{\mathbf{v}}}^{o}|)^{-1}\mathcal{R}\left\{ \text{diag}({\tilde{\mathbf{v}}}^{o*})\mb{M}_\Delta  \right\}\\
\mb{a}&:=|{\tilde{\mathbf{v}}}^{o}|-\mb{A}_Y{\mb{x}}_Y^o-\mb{A}_\Delta{\mb{x}}_\Delta^o
\end{align*}
Using Kirchhoff's laws, we can further derive matrices $\mb{B}, \mb{D}$ and vectors $\mb{b},\mb{d}$. See \cite{linear_2} for a detailed description.
}

\section{Proof of Proposition 1}
{
{
Suppose that $\left\{P_{0,t}^{\vee}\right\}_{t\in\mathcal{T}}$  and $\left\{P_{0,t}^{\wedge}\right\}_{t\in\mathcal{T}}$} are the lower and upper aggregate power trajectories that respect the constraints (\ref{ic}) (\ref{jc}), which are associated with the operational trajectories  $\left\{\mathbf{x}_{t}^{\vee}\right\}_{t\in\mathcal{T}}$ and $\left\{\mathbf{x}_{t}^{\wedge}\right\}_{t\in\mathcal{T}}$ respectively. For any given aggregate power trajectory $\left\{P_{0,t}^o\right\}_{t\in\mathcal{T}}$ that satisfies $ P_{0,t}^{\vee} \leq  P_{0,t}^o \leq P_{0,t}^{\wedge}$ for all $t\in\mathcal{T}$, we define the auxiliary coefficient $\lambda_t\in\left[0,1\right]$ as
$$\lambda_t=\frac{P_{0,t}^{\wedge}-P_{0,t}^{o}}{P_{0,t}^{\wedge}-P_{0,t}^{\vee}}$$
hence $P_{0,t}^o=\lambda_t P_{0,t}^{\vee} +(1-\lambda_t) P_{0,t}^{\wedge}$. Then we claim that $\left\{\mathbf{x}_t^o\right\}_{t\in\mathcal{T}}$ defined by
$$\mathbf{x}_t^o:=\lambda_t\mathbf{x}^{\vee}_t+(1-\lambda_t)\mathbf{x}^{\wedge}_t$$
is a feasible disaggregation solution for $\left\{P_{0,t}^o\right\}_{t\in\mathcal{T}}$, which satisfies constraints (\ref{cf}) with $P_{0,t}^o= \mb{1}_3^\top\mb{D}\mb{x}_t^o+\mb{1}_3^\top\mb{d}_t$. 

Firstly, we show that 
\begin{align*}
\begin{split}
P_{0,t}^o=& (1-\lambda_t)\left( \mb{1}_3^\top\mb{D}\cdot\mb{x}_t^\wedge+\mb{1}_3^\top\mb{d}_t \right)   \\&+ \lambda_t\left( \mb{1}_3^\top\mb{D}\cdot\mb{x}_t^\vee+\mb{1}_3^\top\mb{d}_t \right)
=\mb{1}_3^\top\mb{D}\cdot\mb{x}_t^o+\mb{1}_3^\top\mb{d}_t
\end{split}
\end{align*}

Then from the linear power flow model (\ref{lp1}), $\bv_t$ and $\bi_{L,t}$ are affine functions on $\bx_t$. Thus $\bv_t$ and $\bi_{L,t}$ can be eliminated, and $\mb{g}_t\left(\mb{x}_t,\mb{v}_t,\mb{i}_{L,t}\right)$ can be equivalently rewritten as convex function $\hat{\mb{g}}_t\left(\mb{x}_t\right)$.
Due to the convexity of $\hat{\mb{g}}_t(\cdot)$, we have 
\begin{align*}
\begin{split}
\hat{\mb{g}}_t\left(\mathbf{x}_t^o\right)&=\hat{\mb{g}}_t\left(\lambda_t\mathbf{x}^{\vee}_t+(1-\lambda_t)\mathbf{x}^{\wedge}_t\right) \\
&\leq \lambda_t\hat{\mb{g}}_t\left(\mathbf{x}^{\vee}_t\right)+(1-\lambda_t)\hat{\mb{g}}_t\left(\mathbf{x}^{\wedge}_t\right)\leq \mathbf{0}
\end{split}
\end{align*}
which shows that $\mathbf{x}_t^o$ satisfies time-decoupled constraint (\ref{td}).

Next, we verify that  $\left\{\mathbf{x}_t^o\right\}_{t\in\mathcal{T}}$ also satisfies the time-coupled constraint (\ref{tc}), which involves SOC limits (\ref{dyn}) (\ref{soe}) and the HAVC temperature limits (\ref{temp}) (\ref{comt}). For the SOC constraints (\ref{dyn}) (\ref{soe}), we can reformulate them equivalently as  
\begin{align}\label{nbt}
\frac{\kappa_i^t E_{i,0}-\bar{E}_{i}}{\Delta t}   \leq \sum_{\tau=1}^t\left( \kappa_i^{t-\tau} P_{i,\tau}^e\right)  \leq \frac{\kappa_i^tE_{i,0}-\underline{E}_i}{\Delta t} \ \ \, t\in\mathcal{T}
\end{align}
 Since $\left\{\mathbf{x}_{t}^{\vee}\right\}_{t\in\mathcal{T}}$ and $\left\{\mathbf{x}_{t}^{\wedge}\right\}_{t\in\mathcal{T}}$ are feasible trajectories, their associated ES power outputs $\left\{P_{i,t}^{e,\vee}\right\}_{t\in\mathcal{T}}$ and $\left\{P_{i,t}^{e,\wedge}\right\}_{t\in\mathcal{T}}$ both satisfy constraint (\ref{nbt}).

Therefore by constraint (\ref{jc_bat}), we have
	\begin{align*}
\sum_{\tau=1}^t \left( \kappa_i^{t-\tau} P_{i,\tau}^{e,o}\right) &=\sum_{\tau=1}^t \left[\kappa_i^{t-\tau}\left(  \lambda_\tau P_{i,\tau}^{e,\vee }+(1-\lambda_\tau)P_{i,\tau}^{e,\wedge }\right)\right]\\
	&=\sum_{\tau=1}^t \left[ \kappa_i^{t-\tau}\left(  P_{i,\tau}^{e,\wedge }+\lambda_\tau(P_{i,\tau}^{e,\vee }-P_{i,\tau}^{e,\wedge })\right)\right]\\
	&\geq \sum_{\tau=1}^t \left(  \kappa_i^{t-\tau}P_{i,\tau}^{e,\wedge }\right) \geq  \frac{\kappa_i^t E_{i,0}-\bar{E}_{i}}{\Delta t}
	\end{align*}
	and
		\begin{align*}
\sum_{\tau=1}^t \left( \kappa_i^{t-\tau} P_{i,\tau}^{e,o}\right) 
	&=\sum_{\tau=1}^t [\kappa_i^{t-\tau}( P_{i,\tau}^{e,\vee }-(1-\lambda_\tau)(P_{i,\tau}^{e,\vee }-P_{i,\tau}^{e,\wedge }))]\\
	&\leq \sum_{\tau=1}^t \left( \kappa_i^{t-\tau}P_{i,\tau}^{e,\vee }\right) \leq  \frac{\kappa_i^tE_{i,0}-\underline{E}_i}{\Delta t}
	\end{align*}
which shows that $\left\{\mathbf{x}_t^o\right\}_{t\in\mathcal{T}}$ satisfies the SOC limit constraints (\ref{dyn}) (\ref{soe}). Using a similar argument and constraint (\ref{jc_h}), we prove that $\left\{\mathbf{x}_t^o\right\}_{t\in\mathcal{T}}$ also satisfies the HVAC temperature limit constraints (\ref{temp}) (\ref{comt}).

In this way, we prove that there exists a feasible disaggregation solution $\left\{\mathbf{x}_t^o\right\}_{t\in\mathcal{T}}$  for the given trajectory $\left\{P_{0,t}^o\right\}_{t\in\mathcal{T}}$.
}

%

\vskip -4pt plus -1fil

\begin{IEEEbiography}[{\includegraphics[width=1in,height=1.25in,clip,keepaspectratio]{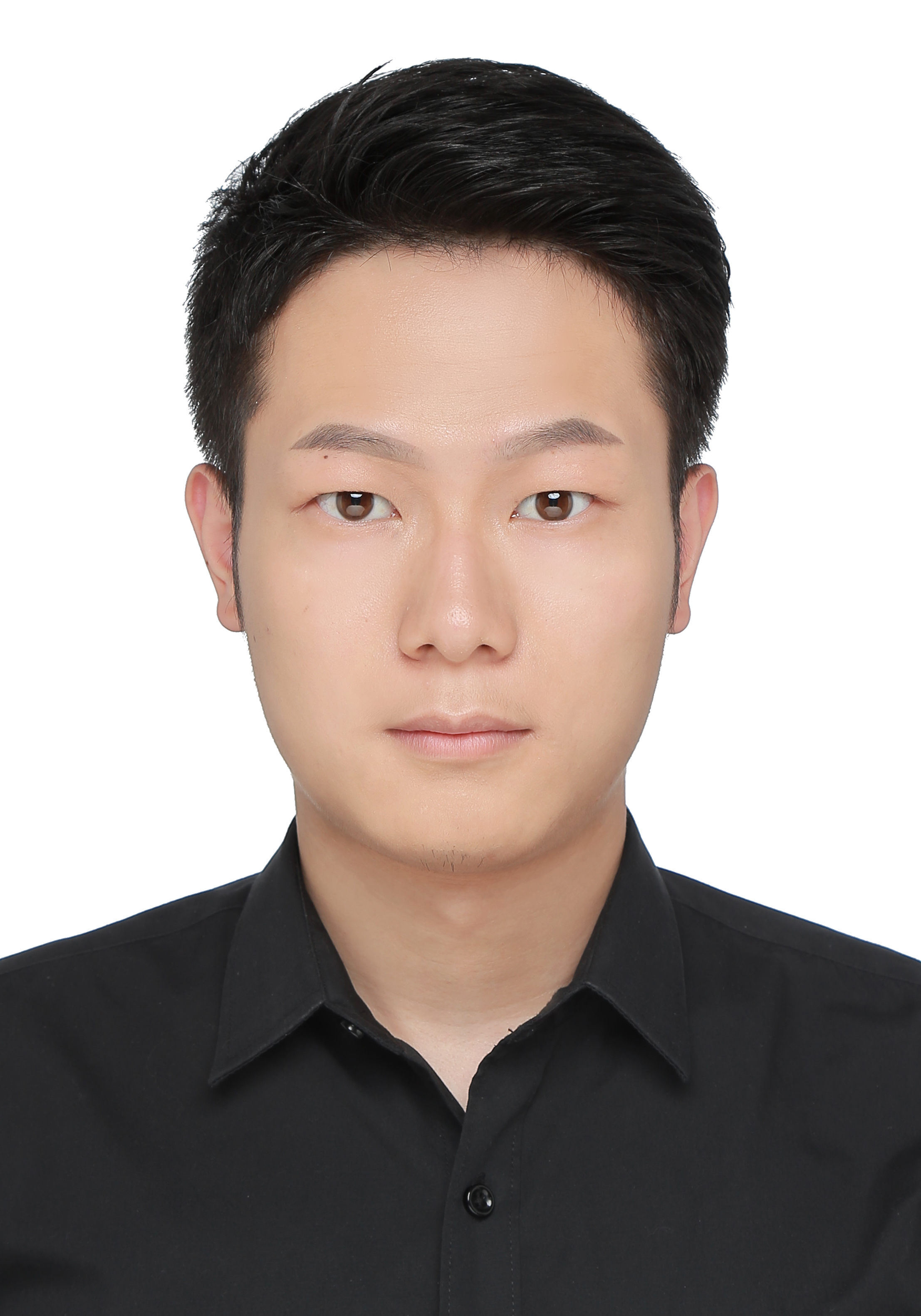}}]{Xin Chen}
received his double B.S. degrees in engineering physics and economics from Tsinghua University, Beijing, China in 2015, and the master degree in electrical engineering from Tsinghua University, Beijing, China in 2017. He is 
currently pursuing the Ph.D. degree in electrical engineering at Harvard University, USA. 
He received the Best Conference Paper Award in IEEE PES General Meeting 2016, and he was a Best Student Paper Award Finalist in the IEEE Conference on Control Technology and Applications (CCTA) 2018. His research interests focus on 
distributed optimization and control of networked systems, with emphasis on power systems.
\end{IEEEbiography}

\vskip 0pt plus -1fil

\begin{IEEEbiography}[{\includegraphics[width=1in,height=1.25in,clip,keepaspectratio]{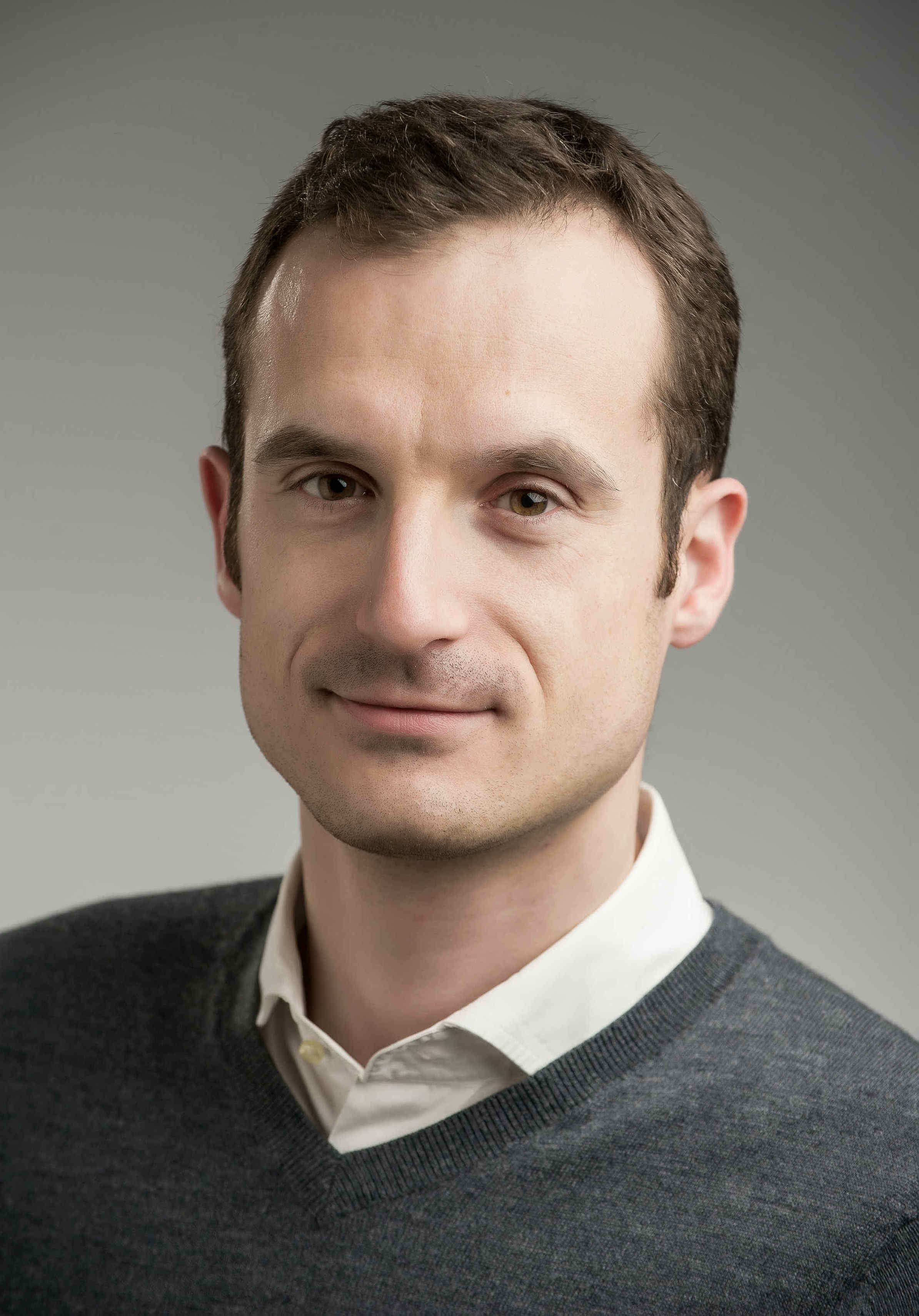}}]{Emiliano Dall'Anese}
	(S'08-M'11) received the Laurea Triennale (B.Sc Degree) and the Laurea Specialistica (M.Sc Degree) in Telecommunications Engineering from the University of Padova, Italy, in 2005 and 2007, respectively, and the Ph.D. in Information Engineering from the Department of Information Engineering, University of Padova, Italy, in 2011. From January 2009 to September 2010, he was a visiting scholar at the Department of Electrical and Computer Engineering, University of Minnesota, USA. From January 2011 to November 2014, he was a Postdoctoral Associate at the Department of Electrical and Computer Engineering and Digital Technology Center of the University of Minnesota, USA. From December 2014 to August 2018, he was a Senior Researcher at the National Renewable Energy Laboratory, Golden, CO, USA. Since August 2018 he has been an Assistant Professor within the department of Electrical, Computer, and Energy Engineering at the University of Colorado Boulder. His research interests focus on optimization, decision systems, and statistical learning, with applications to networked systems and cyber-physical systems.   
\end{IEEEbiography}

\vskip 0pt plus -1fil

\begin{IEEEbiography}[{\includegraphics[width=1in,height=1.25in,clip,keepaspectratio]{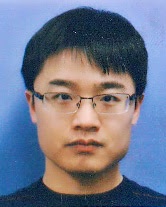}}]{Changhong Zhao}
	(S'12-M'15) is an Assistant Professor with the Department of Information Engineering, the Chinese University of Hong Kong. He received the B.Eng. degree in automation from Tsinghua University, Beijing, China, in 2010, and the PhD degree in electrical engineering from California Institute of Technology in 2016. He was awarded the Caltech Demetriades PhD Thesis Prize and Wilts PhD Thesis Prize. From 2016 to 2019, he was a researcher at the National Renewable Energy Laboratory, Golden, CO, USA. His research interests include distributed control of networked systems and cyber-physical systems, power system dynamics and stability, and optimization of multi-energy systems and other smart-city infrastructure.
\end{IEEEbiography}

\vskip 0pt plus -1fil

\begin{IEEEbiography}[{\includegraphics[width=1in,height=1.25in,clip,keepaspectratio]{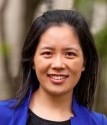}}]{Na Li}
 received her B.S. degree in mathematics and applied mathematics from Zhejiang University in
	China and her PhD degree in Control and Dynamical Systems from the California Institute of Technology
	in 2013. She is an Associate Professor in the School of Engineering and Applied Sciences at Harvard
	University. She was a postdoctoral associate of the Laboratory for Information and Decision Systems at
	Massachusetts Institute of Technology. Her research lies in the design, analysis, optimization,
	and control of distributed network systems, with particular applications to cyber-physical network systems. She received NSF CAREER Award in 2016, AFOSR Young Investigator Award in 2017,
	ONR Young Investigator Award in 2019 among others. 
\end{IEEEbiography}

	
\end{document}